\documentclass[aps,prb,citeautoscript,onecolumn,superscriptaddress,notitlepage]{revtex4-1}
\usepackage[T1]{fontenc} 
\usepackage{physics,amsmath,amssymb,amsthm,bm,setspace,graphicx}
\usepackage{multirow,array}
\usepackage[table]{xcolor}

\begin{document}

\title{On Stretching, Bending, Shearing and Twisting of Actin Filaments II: 
\break Multi-Resolution Modelling}

\author{Ravinda S. Gunaratne}
\affiliation{{\rm Mathematical Institute, University of Oxford, Radcliffe Observatory Quarter, Woodstock Road, Oxford, OX2 6GG, United Kingdom}}
\author{Carlos Floyd}
\affiliation{{\rm Department of Chemistry \& Biochemistry,
Institute for Physical Science and Technology,
University of Maryland,
College Park, MD 20742, USA; \\ e-mails: gpapoian@umd.edu, erban@maths.ox.ac.uk}}
\author{Haoran Ni}
\affiliation{{\rm Department of Chemistry \& Biochemistry,
Institute for Physical Science and Technology,
University of Maryland,
College Park, MD 20742, USA; \\ e-mails: gpapoian@umd.edu, erban@maths.ox.ac.uk}}
\author{Garegin A. Papoian}
\affiliation{{\rm Department of Chemistry \& Biochemistry,
Institute for Physical Science and Technology,
University of Maryland,
College Park, MD 20742, USA; \\ e-mails: gpapoian@umd.edu, erban@maths.ox.ac.uk}} 
\author{Radek Erban}
\affiliation{{\rm Mathematical Institute, University of Oxford, Radcliffe Observatory Quarter, Woodstock Road, Oxford, OX2 6GG, United Kingdom}}

\begin{abstract}
\noindent
{\bf Abstract.}
We present a multi-resolution methodology for modelling F-actin filaments. It provides detailed microscopic information at the level of individual monomers at a lower computational cost by replacing the monomer-based model in parts of the simulated filament by a rod-based macroscopic model. In the monomer-based description, G-actin is represented by ellipsoids bound at the surface in a double helical configuration to form F-actin. The rod-based model is coarser, in which F-actin is described using a Cosserat model, as seen in the preceding paper~\cite{Floyd:2022:SBSI}. The multi-resolution methodology is illustrated using three case studies, designed to test the properties of F-actin under stretching, bending, shearing and twisting. The methodology is especially suited for situations where filaments are subject to bending deformations. We investigate the limitations of using the standard Cosserat model to capture the complete torsional behaviour of F-actin, presenting its extensions which account for curvature dependent rigidities and a twist-stretch coupling to improve accuracy of the overall multi-resolution scheme. 
\end{abstract}

\maketitle

\section{Introduction}

\noindent
Actin is a globular protein that forms microfilaments which can be found in all eukaryotic cells. It exists either as free G-actin monomers, or in F-actin filaments with the monomers forming thin, stiff double helical strands. Actin filaments are found in the cellular cytoskeleton in structures like filopodia and lamellipodia, playing a role in many important processes, such as cell migration and muscle contraction via actomyosin networks. In these settings we find actin interacting with numerous binding proteins and crosslinkers to give the networks the biological structures required to perform their function~\cite{Lodish:2021:MCB}.

As a result of its importance and varied role in these situations, the behaviour of actin has been investigated at many biological length and time scales. Detailed all-atom molecular dynamics (MD) simulations can be parametrised using data obtained via X-ray crystallography~\cite{Dominguez2011ActinFunction} or cryogenic electron microscopy ~\cite{Merino2018StructuralCryo-EM,Chou2019MechanismNucleotides} available in repositories such as the RCSB Protein Data Bank \cite{Berman2000TheBank}. These simulations capture the fine chemical bonding structure of actin and bound ligands, both in monomer and filamentous forms~\cite{Pfaendtner2010StructureFilament,Shamloo2018NanomechanicsSimulation,Tanaka2018StructuralDisassembly,Narita2006StructuralStudy,Oda2009TheTransition}, though this detail comes at a high computational cost. Atomistic simulations can be used to investigate the physical properties of short F-actin sections~\cite{Chu2005AllosteryAnalysis}, but it is infeasible to use this level of detail for large systems of biological interest.

Coarse-grained models of actin monomers offer a significant reduction in computational cost when compared to all-atom simulations. Actin monomers, either in free or bound filament form, can be represented as spheres or ellipsoids~\cite{Dobramysl2016StericFilopodia,Schramm2017ActinDissociation,Schramm2019PlasticFilaments}, in a more detailed 4-bead description~\cite{Chu2005AllosteryAnalysis,Chu2006Coarse-grainedSimulations}, or an even greater level of detail that begins to bridge the gap in resolution to all-atom simulations~\cite{Horan2019InsightsModel,Katkar2018InsightsFilaments}. Parameters for these coarse-grained models can be set either from experimental data or by inferring them, using various techniques~\cite{Chu2006Coarse-grainedSimulations,Savelyev2010ChemicallyDNA}, from all-atom MD simulation data.~\cite{Aydin2018MultiscaleProteins,Yamaoka2012MultiscaleCytoskeleton}. In this paper, we will use a monomer-based description of actin, which was developed by De La Cruz et al.~\cite{DeLaCruz2010OriginFilaments}, who parametrised it using atomistic MD simulations. This model will form our monomer-based description of actin introduced in  Section~\ref{secmicromodel}.

While coarse-grained monomer-based models offer an efficiency improvement over all-atom MD simulations, they can still be hard to scale up to more biologically complex systems. Multi-resolution modelling techniques can overcome this by describing less important parts of the simulated system using coarser and more efficient models. Examples include coupling atomistically detailed regions for an active site of a protein to coarser representations of the remainder of a protein~\cite{Fogarty2016ASite}, or using less detailed descriptions of the parts of a DNA molecule, which are relatively far from the DNA-protein interaction site of interest~\cite{Rolls2016VaryingDynamics,Rolls2018Multi-resolutionInteractions}. Multi-resolution modelling has also been used in the literature to coarse-grain the description of the solvent in the parts of computational domain, which is relatively far from the biological structure of interest~\cite{Zavadlav2015AdaptiveSolution,Chapman,Praprotnik2007ASimulation}, and for replacing the explicit solvent representation in parts of the domain by a stochastic coarse-grained model, which describes the solvent implicitly~\cite{Erban2014FromDynamics,Gunaratne2019Multi-resolutionInteractions,Erban2016CouplingDynamics}. To obtain multi-resolution methods with mutually consistent microscopic and macroscopic models, the parameters of such stochastic coarse-grained models can be inferred from all-atom MD simulations~\cite{Erban2016CouplingDynamics,Erban2019Coarse-grainingDistributions}, by estimating effective force distributions on coarse-grained sites~\cite{Wang:2009:EFC,Joshi:2020:RAC,Utterson:2022:SMF}. In this way, both microscopic and macroscopic models consistently provide quantitatively the same information at the macroscopic level, while the monomer-based model provides an additional level of detail in the parts of the simulated system which is of interest to a modeller. In addition to this `bottom-up' approach, macroscopic models of actin filaments can also be developed by parametrising them against the available experimental data~\cite{Popov2016MEDYAN:Networks,Floyd2019QuantifyingNetworks,MacKintosh1995ElasticityNetworks,Satcher1996TheoreticalCytoskeleton}. 

In this paper, our macroscopic model of an actin filament will utilise a Cosserat description developed by Gazzola et al.~\cite{Gazzola2018ForwardFilaments}, which we introduce in Section~\ref{secmacromodel}. It considers torsional, bending, shear and compression forces on the actin filament. The microscopic monomer-based model introduced in Section~\ref{secmicromodel} describes these forces with a higher level of detail, but less efficiently: we will call it the Ellipsoid model in what follows. We show how the Ellipsoid and Cosserat models can be used in a multi-resolution framework, where the monomer-based description is substituted in parts of the filament by the rod-based description to improve overall simulation efficiency. Our multi-resolution coupling strategy is explained in Section~\ref{secmultiactin}, which is followed by three numerical examples in Section~\ref{secnumexamples}. We conclude in Section~\ref{secdiscussion} by discussing biological settings where a multi-resolution model of actin could be particularly helpful.

\section{Ellipsoid and Cosserat models of actin filaments}

In this section, we introduce both Ellipsoid and Cosserat models of an actin filament. The Ellipsoid model describes individual G-actin monomers as ellipsoids and was developed by De La Cruz et al.~\cite{DeLaCruz2010OriginFilaments}, who used a fluctuation matching technique to parametrise the equations of motion using atomistic MD simulation data. The Cosserat model is a generalization of the rod-based filament model developed by Gazzola et al.~\cite{Gazzola2018ForwardFilaments} and we will introduce it in Section~\ref{secmacromodel}.

\subsection{Ellipsoid model}
\label{secmicromodel}

\begin{figure*}[t]
\centering
\includegraphics[width=0.9\textwidth]{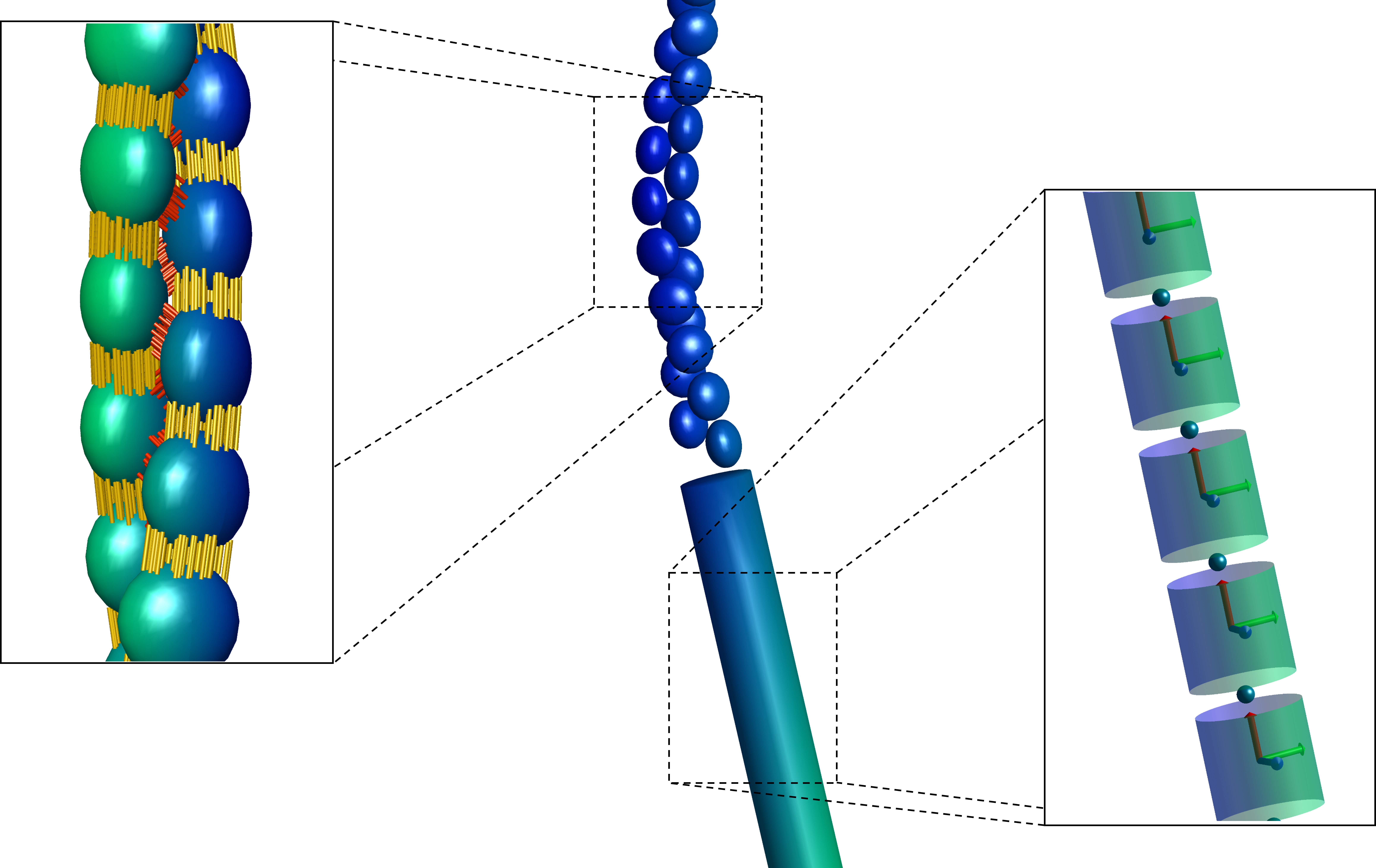}
\caption[A schematic of Ellipsoid and Cosserat models in the multi-resolution framework.]{\label{figure1}
{\it A schematic of the Ellipsoid model described in Section~$\ref{secmicromodel}$ 
(left panel) and the Cosserat model described in Section~$\ref{secmacromodel}$
(right panel) which are coupled in the multi-resolution framework (middle panel).}}
\end{figure*}

The actin filament is constructed out of ellipsoid shaped monomers placed along a double helix, as is schematically shown in the left panel of Figure~\ref{figure1}. The dimensions and spatial positions of these monomers are taken from crystal structures of F-actin filaments~\cite{DeLaCruz2010OriginFilaments}. The centre of the $i^{\mathrm{th}}$ monomer is denoted by $\bm{\Bar{R}}_{i}$, where $i=1,2,\dots,N$, and $N$ is the total number of monomers. As shown in Figure~\ref{figure1}, each monomer is connected by harmonic bonds to its two nearest neighbours up and down the chain, if they exist. That is, the $i^{\mathrm{th}}$ monomer is connected by $n_\parallel$ harmonic bonds to the next and previous monomer in their strand of the double helix,
which are labelled as the $(i+2)^{\mathrm{th}}$ and $(i-2)^{\mathrm{th}}$ monomers. Moreover, the $i^{\mathrm{th}}$ monomer is also connected by $n_\perp$ harmonic bonds to each of the adjacent monomers in the offset strand of the double helix, which are labelled as the $(i+1)^{\mathrm{th}}$ and $(i-1)^{\mathrm{th}}$ monomers.

The monomers are connected to each other by harmonic bonds, with the end points of these placed on the surfaces of the ellipsoids. This allows for both the distance between monomer centres and their relative angular orientations to be constrained. Considering the $k^{{\rm th}}$ bond between the $i^{{\rm th}}$ and $j^{{\rm th}}$ monomers, where $i=1,2,\dots,N$ and $k=1,2,\dots,n_\perp$ for $j=i \pm 1$ or $k=1,2,\dots,n_\parallel$ for $j= i \pm 2$, we denote by $\bm{r}^{k}_{i}$ and $\bm{r}^{k}_{j}$ the end points of the $k^{\mathrm{th}}$ bond on the surface of the $i^{{\rm th}}$ and $j^{{\rm th}}$ monomers, respectively. Then we define the force at $\bm{r}^{k}_{i}$, resulting from the $k^{{\rm th}}$ bond between the $i^{{\rm th}}$ and $j^{{\rm th}}$ monomers, as 
\begin{equation}
\bm{F}_{ij}^k =  S_{ij} \, (\bm{r}^{k}_{j} - \bm{r}^{k}_{i}) \left(1 - \frac{L^k_{ij}}{|\bm{r}^{k}_{j} - \bm{r}^{k}_{i}|}\right),
\qquad
\mbox{for} 
\quad |i-j| = 1,2,
\label{indivforces}
\end{equation}
where the spring constants $S_{ij}$ take different values for lateral and longitudinal bond interactions, namely
\begin{equation*}
S_{ij} 
= 
\begin{cases}
S_\perp/n_\perp, & \text{for $|i-j|=1$},\\
S_\parallel/n_\parallel, & \text{for $|i-j|=2$},
\end{cases}
\end{equation*}
where $S_\parallel$ and $S_\perp$ are the total strains, and $L^k_{ij}$ in equation~(\ref{indivforces}) is the equilibrium length of the $k^{{\rm th}}$ bond between the $i^{{\rm th}}$ and $j^{{\rm th}}$ monomers. 
By summing up forces~(\ref{indivforces}), we can define the overall translational and torsional effects on the $i^{{\rm th}}$ monomer, relative to the centre of this ellipsoid, $\bm{\Bar{R}}_{i}$, as follows:
\begin{equation*}
\bm{\Bar{F}}_{i} = \sum\limits_{j,k} \bm{F}_{ij}^k,
\qquad
\mbox{and}
\qquad
\bm{\Bar{\tau}}_i = \sum\limits_{j,k} (\bm{r}_{i}^k - \bm{\Bar{R}}_{i}) \cross \bm{F}_{ij}^k \,,
\end{equation*}
where we sum over all bonds between the $i^{{\rm th}}$ monomer and each of the monomers that it is bonded to. Equations of motion are written in terms of positions $\bm{r}^{k}_{i}$ and $\bm{\Bar{R}}_{i}$ as follows
\begin{eqnarray}
\frac{\mbox{d}\bm{r}_i^k}{\mbox{d}t}
& \! = \! &
\bm{\omega}_i \times
(\bm{r}_i^k - \bm{\Bar{R}}_{i}) 
 + 
\bm{v}_i \, ,
\hskip 3.45cm
\frac{\mbox{d}\bm{\Bar{R}}_{i}}{\mbox{d}t}
 \, = \, \bm{v}_i \, ,
\hskip 1cm
\label{posmicroeq}
\\
m\frac{\mbox{d}\bm{v}_i}{\mbox{d}t}
& = & \bm{\Bar{F}}_{i} \, ,
\hskip 6.0cm
\bm{I}\frac{\mbox{d}\bm{\omega}_i}{\mbox{d}t}
\, = \, (\bm{I}\bm{\omega}_i) \cross \bm{\omega}_i + 
\bm{\Bar{\tau}}_i \, ,
\hskip 1cm
\label{velmicroeq}
\end{eqnarray}
where $\bm{v}_i$ and $\bm{\omega}_i$ are the velocity and angular velocity of the $i^{{\rm th}}$ monomer, respectively, $m$ is the mass of a monomer, and $\bm{I}$ is the moment of inertia tensor. To discretize equations~\hbox{(\ref{posmicroeq})--(\ref{velmicroeq})}, it is possible to use a number of symplectic integrators. In this paper, we apply the 4$^{{\rm th}}$ order symplectic method presented by Omelyan, Mryglod and Folk~\cite{Omelyan2002OptimizedSystems}. The orientation of the $i^{{\rm th}}$ monomer is characterized by the matrix $\bm Q_i$, for $i=1,2,\dots,N$, whose columns are three orthonormal vectors $\bm{d}_{1,i}$, $\bm{d}_{2,i}$, and $\bm{d}_{3,i}$, where $\bm{d}_{3,i}$ is the direction of the long filament axis. By rotating this basis at each timestep, rather than all individual bond endpoints, we can express their position at a given time $t$ using
\begin{equation*}
\bm{r}_i^k(t) = \bm{Q}_i(t)\,\bm{Q}^T_i(0)     
\left(\bm{r}_i^k(0) - \Bar{\bm{R}}_i(0) \right) 
+ \Bar{\bm{R}}_i(t) \, . 
\end{equation*}
The time evolution of $\bm{d}_{j,i}$ is for $j=1,2,3$ given by
\begin{equation*}
 \frac{\mbox{d}\bm{d}_{j,i}}{\mbox{d}t} 
=
(\bm{Q}_i\,\bm{\omega}_{i,\mathcal{L}}) \cross \bm{d}_{j,i}  \,  , 
\end{equation*}
where each respective element of $\bm{\omega}_{i,\mathcal{L}}$ is the angular velocity around the $\bm{d}_{j,i}$ vectors that make up the orientation basis $\bm{Q}_i$. This reduces numerical errors, in addition to being useful when defining our multi-resolution coupling in Section~\ref{secmultiactin}.

\subsection{Cosserat model}
\label{secmacromodel}

In the work by Gazzola et al.~\cite{Gazzola2018ForwardFilaments} a rod-based filament model is presented, with torsional, bending, shear and compression forces accounted for. The equations of motion are derived by first considering an inextensible and unshearable Kirchoff-Love rod model \cite{Landau1986TheoryElasticity}, then extending this to a Cosserat model of stretchable and shearable filaments.
A filament is described by its centre-line $\bm{r}(s,t)$ with velocity $\bm{v}(s,t)$, where $t$ is time and $s$ is the centre-line arc coordinate. We also define an orientated frame of reference given by vectors $\bm{d}_{1}(s,t)$, $\bm{d}_{2}(s,t)$, and $\bm{d}_{3}(s,t)$, which form an orthonormal basis of unit vectors, and the corresponding matrix 
$\bm{Q}(s,t) = 
\big(
\bm{d}_{1}(s,t),\bm{d}_{2}(s,t),\bm{d}_{3}(s,t)
\big)$.
We define the laboratory and body-convected frames as
$\bm{x} = \widehat{x_1} \,\widehat{\bm{d}}_1 + \widehat{x_2} \,\widehat{\bm{d}}_2 + \widehat{x_3} \,\widehat{\bm{d}}_3$
and
$\bm{x}_{\mathcal{L}} = \bm{Q}^T\bm{x} = x_1 \bm{d}_{1} + x_2 \bm{d}_{2} + x_3 \bm{d}_{3},$
respectively, where the laboratory reference basis is denoted $\big\{ \widehat{\bm{d}}_1,\widehat{\bm{d}}_2,\widehat{\bm{d}}_3\big\}$ as in the preceding paper~\cite{Floyd:2022:SBSI}. The variation of matrix $\bm{Q}$ in space (along the filament) and in time can be used to formally consider the curvature and angular momentum. We define the rod angular velocity and generalised curvature respectively as 
\begin{equation*}
\bm{\kappa} 
= 
\text{ax}
\left[\bm{Q}^T \,\pdv{\bm{Q}}{s} \right],
\qquad
\bm{\omega} = \mathrm{ax}\left[\bm{Q}^T \,\pdv{\bm{Q}}{t} \right],
\end{equation*}
where the $\text{ax}$ operation returns the pseudovector associated with the skew-symmetric matrix; see the footnote in the accompanying paper~\cite{Floyd:2022:SBSI} for exact definition. 

Following Gazzola et al.~\cite{Gazzola2018ForwardFilaments}, we split the filament into $n$ segments, with each segment defined by their endpoints $\bm{r}_i(t)$ for $i = 1,2,\dots,n+1$. Additional quantities associated with the endpoints (their velocity $\bm{v}_{i}(t)$ and applied force $\bm{F}_{i}(t)$) have $n + 1$ elements, while discretized quantities associated with the segments have $n$ elements. We define the segment vector, deviation from its rest length (i.e. the local stretching/compression) and normal tangent as
\begin{equation*}
\bm{l}_i = 
\bm{r}_{i+1} - \bm{r}_{i} \,, 
\qquad 
e_i = \frac{|\bm{l}_i|}{|\hat{\bm{l}}_i|} \,, 
\qquad \bm{t}_i = \frac{\bm{l}_i}{|\bm{l}_i|},
\qquad
\mbox{for} \;\; i=1,2,\dots,n,
\end{equation*}
where $|\hat{\bm{l}}_i|$ is the rest length of the segment, and for all following definitions the hat notation indicates a rest, un-stretched quantity. In the continuum setting all quantities can be defined point-wise, whereas in a discrete setting some quantities (such as $\hat{\bm{\kappa}}_\mathcal{L}$) are naturally expressed over an integrated domain $\mathcal{D}$ along the filament,\cite{Audoly2013AThreads} with the Voronoi region, and associated compression factor defined as
$\mathcal{D}_i = (|\bm{l}_{i+1}| + |\bm{l}_i|)/2$ and $\mathcal{E}_i = \mathcal{D}_i/\hat{\mathcal{D}_i}.$
As the generalised curvature expresses a rotation per unit length, the quantity $\hat{\mathcal{D}}_i \hat{\bm{\kappa}}_\mathcal{L}^{i}$ expresses the rotation that transforms a material frame to its neighbouring frame over the segment, therefore
$\exp\!\big(\hat{\mathcal{D}}_i \hat{\bm{\kappa}}_\mathcal{L}^{i} \big) \,
\bm{Q}^T_i = \bm{Q}^T_{i+1}.$
Having set out these quantities, we can now define the discrete shear and curvature vectors for each segment as
\begin{equation}
\bm{\sigma}_{\mathcal{L}}^i = \bm{Q}^T_i \left( e_i \bm{t}_i - \bm{d}_{3,i}\right),
\qquad
\mbox{and}
\qquad
\hat{\bm{\kappa}}_\mathcal{L}^i = \frac{\log(\bm{Q}^T_{i+1}\bm{Q}_{i})}{\hat{\mathcal{D}}_i}.
\label{curvvector}
\end{equation}
We can finally write the full spatially discretised equations of motion 
\begin{eqnarray*}
 m_i \pdv{\bm{v}_i}{t} &=& \Delta^h\left(\frac{\bm{Q}_i \hat{\bm{S}}_i\bm{\sigma}^i_{\mathcal{L}}}{e_i}\right) + \bm{F}_i, \\
 \left(\frac{\hat{\bm{I}_i}}{e_i}\right) \pdv{\bm{\omega}^i_{\mathcal{L}}}{t} &=& \Delta^h\left(\frac{\hat{\bm{B}}_i \hat{\bm{\kappa}}^i_{\mathcal{L}}}{\mathcal{E}_i^3}\right) + \mathcal{A}^h\left(\frac{\hat{\bm{\kappa}}^i_{\mathcal{L}} \cross \hat{\bm{B}}_i \hat{\bm{\kappa}}^i_{\mathcal{L}}}{\mathcal{E}_i^3}\hat{\mathcal{D}}_i\right) + \left(\bm{Q}^T_i \bm{t}_i \cross \hat{\bm{S}}_i\bm{\sigma}^i_{\mathcal{L}}\right) |\hat{\bm{l}}_i| \nonumber\\ 
 &&+ \frac{ \hat{\bm{I}}_i\bm{\omega}^i_{\mathcal{L}}}{e_i^2}\pdv{e_i}{t} + \left(\frac{ \hat{\bm{I}}_i\bm{\omega}^i_{\mathcal{L}}}{e_i}\right) \cross \bm{\omega}^i_{\mathcal{L}} + \bm{C}^i_{\mathcal{L}},
\end{eqnarray*}
where $\hat{\bm{I}}_i$ is the inertia tensor, and $\hat{\bm{B}}$ and $\hat{\bm{S}}$ are the bend/twist and shear/stretch rigidity matrices, respectively. The  difference and trapezoidal quadrature operators used are defined as
\begin{equation}
\Delta^h(\bm{x}_{i}) 
= 
\begin{cases}
\bm{x}_1, & \text{if $j=1$},\\
\bm{x}_j - \bm{x}_{j-1}, & \text{if $1<j\leq N$},\\
-\bm{x}_N, & \text{if $j=N+1$}.
\end{cases}
\qquad \quad
\mathcal{A}^h(\bm{x}_{i}) = 
\begin{cases}
\bm{x}_1/2, & \text{if $j=1$},\\
(\bm{x}_j + \bm{x}_{j-1})/2, 
& \text{if $1<j\leq N$},\\
\bm{x}_N/2, & \text{if $j=N+1$}.
\end{cases}
\label{trapoper}
\end{equation}
When calculating potential energy, we integrate over the length of the filament for both the twist/bend and shear/stretch terms
\begin{equation}\label{eq:cosserat_energy}
    E = \frac{1}{2}\int\limits_{0}^{L} \bm{\kappa}_{\mathcal{L}}^T \bm{B}\bm{\kappa}_{\mathcal{L}}\, \mathrm{d}s + \frac{1}{2}\int\limits_{0}^{L} \bm{\sigma}_{\mathcal{L}}^T \bm{S}\bm{\sigma}_{\mathcal{L}}\, \mathrm{d}s . 
\end{equation}
Entries of the rigidities matrices are usually constant, though it is possible to formulate a generalised model, where $\hat{\bm{B}}(\bm{\kappa}_{\mathcal{L}})$ or $\hat{\bm{S}}(\bm{\sigma}_{\mathcal{L}})$. This is particularly relevant when modelling actin, due to its asymmetric torsional properties. By allowing the twisting rigidity $B_3(\kappa_{\mathcal{L},3})$ to be curvature dependent, the accuracy of the Cosserat model can be improved to better match the properties of our microscopic Ellipsoid model. This extension will be studied in more detail in Section~\ref{sectwist}. For this work we use the same 4$^\mathrm{th}$ order symplectic integrator~\cite{Omelyan2002OptimizedSystems} as employed for the Ellipsoid model.

\section{Multi-resolution modelling framework}
\label{secmultiactin}

To couple the Ellipsoid model of actin filament (described in Section~\ref{secmicromodel}) to the Cosserat model (described in Section~\ref{secmacromodel}), we must find a way to mediate transfer of forces between models at boundary regions. This involves projecting forces resulting from each of the harmonic bond anchor sites on the monomers in the coupling region to forces acting solely on the angular and translational components of one (or more) rod segments of the Cosserat model. 
\begin{figure*}
\centering
\includegraphics[width=0.9\textwidth]{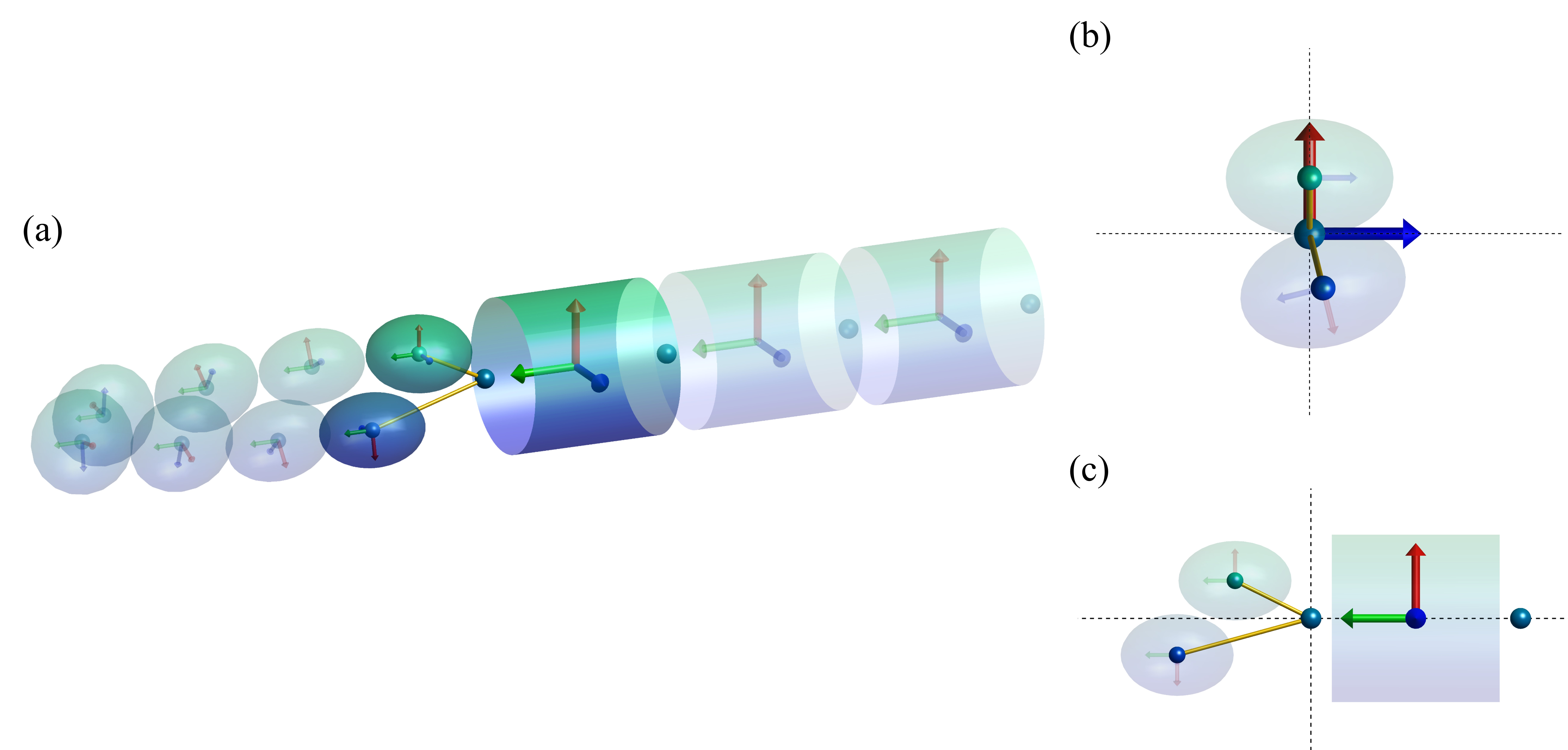}
\caption{\label{figure2} (a) {\it Illustration of multi-resolution coupling region. The final Cosserat model segment is bound to the final two monomers in the Ellipsoid model. This is achieved by attaching the final segment to the ellipsoid centres with bonds (yellow) that have the same material properties dictated by the model parameters of the Cosserat model.} \hfill\break (b) {\it View along the centre line.} (c) {\it Side view.}}
\end{figure*}
Both the Ellipsoid and Cosserat models have orientation matrices for monomers and rods, respectively, with their angular velocities defined in their local frame. To capture twisting behaviour, we must define forces that are dependent on orientation in addition to the endpoint and ellipsoid centre positions. Below we present a method to achieve this in a manner consistent with the formulation of the Cosserat model. The schematic of the multi-resolution coupling region is shown in Figure~\ref{figure2}.

The rod-based part of the filament is described by the centre-line $\bm{r}(s,t)$, where $s \in [0,1]$, and we identify $s=0$ with the endpoint, which is attached to the monomer-based Ellipsoid model, i.e. $\bm{r}(0,t)$ is the endpoint of the final segment of the Cosserat model and $\bm{Q}(0,t)$ is the corresponding orientation basis. Labelling the monomer closest to $\bm{r}(0,t)$ as the monomer~$1$, we intend to attach the rod-based part of the filament to the monomer with its centre denoted by $\bm{\Bar{R}}_{1}$ and its orientation basis by $\bm{Q}_{1}$. To do this, we create a new bond between $\bm{r}(0,t)$ and $\bm{\Bar{R}}_{1}$. We define an orientation basis for the coupling $\bm{Q}_{\mathrm{MR}}$ and the coupling direction vector, extension, tangent, and strain in a manner consistent with the Cosserat model, by
\begin{equation*}
\bm{l}_{\mathrm{MR}} 
= 
\bm{r}(0,t)
- 
\bm{\Bar{R}}_{1} \, ,
\quad 
e_{\mathrm{MR}} 
= 
\frac{|\bm{l}_{\mathrm{MR}}|}{|\hat{\bm{l}}_{\mathrm{MR}}|} 
\, , \quad 
\bm{t}_{\mathrm{MR}} = \frac{\bm{l}_{\mathrm{MR}}}{|\bm{l}_{\mathrm{MR}}|} \, ,
\quad
\bm{\sigma}_{\mathrm{MR}} = \bm{Q}^T_{\mathrm{MR}} \left( e_{\mathrm{MR}} \bm{t}_{\mathrm{MR}} - \bm{Q}_{\mathrm{MR}} \bm{d}_{\mathrm{MR}}\right),
\end{equation*}
where subscript MR corresponds in our notation to the multi-resolution coupling region. The key difference occurs in the definition of the strain, where, rather than using the basis vector which is parallel to the backbone $\bm{d}_3$, we instead use the rest vector direction between $\bm{r}(0,t)$ and $\bm{\Bar{R}}_{1}$ given by $\bm{d}_{\mathrm{MR}} = \hat{\bm{l}}_{\mathrm{MR}}/|\hat{\bm{l}}_{\mathrm{MR}}|$, and rotate this vector to the current orientation of the coupling bond. This allows us to initialise the model such that the orientation basis is aligned with the filament direction, even though the coupling bond is not, so in the case of a straight, untwisted filament, we have $\bm{Q}(0,t) = \bm{Q}_{1} = \bm{Q}_{\mathrm{MR}}.$ The advantage of proceeding in this manner is that it allows us to define a strain matrix $\bm{S}$ in a manner consistent with the Cosserat model, with the correct strains both axially and radially. First, we define the linear force experienced by the final segment endpoint and ellipsoid centre as a result of the coupling's shearing and stretching properties
\begin{equation*}
    \bm{F}_{\mathrm{Cos}} = \frac{\bm{Q}_{\mathrm{MR}} \hat{\bm{S}}\bm{\sigma}_{\mathrm{MR}}}{e_{\mathrm{MR}}} \quad , \quad \bm{F}_{\mathrm{Ell}} = - \bm{F}_{\mathrm{Cos}},
\end{equation*}
with quantities labelled with subscripts MR, Cos, and Ell corresponding to the multi-resolution coupling region, Cosserat model, and Ellipsoid model, respectively. While this force keeps the segment and ellipsoids in the coupling in the correct spatial positions, we introduce an angular constraint to ensure the ellipsoid monomers maintain the correct orientation relative to the final Cosserat segment. First, we define the rest lengths of the Voronoi regions and curvatures of the Cosserat segment relative to the coupling, and the coupling relatives to the ellipsoid  
\begin{equation*}
    \hat{D}_{\mathrm{Ell}} = \frac{|\widehat{\bm{d}}_3\cdot\bm{Q}_{\mathrm{MR}}^T\hat{\bm{l}}_{\mathrm{MR}}|}{2} \, , \quad
    \hat{D}_{\mathrm{Cos}} = \frac{|\hat{\bm{l}}_i|}{2} +  \frac{|\widehat{\bm{d}}_3\cdot \bm{Q}_{\mathrm{MR}}^T\hat{\bm{l}}_{\mathrm{MR}}|}{2} \, , \quad
    \mathcal{E}_{\mathrm{Cos}} = \frac{D_{\mathrm{Cos}}}{\hat{D}_{\mathrm{Cos}}} \, , \quad
    \mathcal{E}_{\mathrm{Ell}} = \frac{D_{\mathrm{Ell}}}{\hat{D}_{\mathrm{Ell}}},
\end{equation*}
\begin{equation*}
    \hat{\bm{\kappa}}_{\mathrm{Ell}} = \frac{\log( \bm{Q}^T_1\bm{Q}_{\mathrm{MR}})}{\hat{D}_{\mathrm{Ell}}} \, , \quad
    \hat{\bm{\kappa}}_{\mathrm{Cos}} = \frac{\log(\bm{Q}^T_{\mathrm{MR}} \bm{Q}(0,t))}{\hat{D}_{\mathrm{Cos}}}\, ,
\end{equation*}
where $|\hat{\bm{l}}_i|$ is the rest length of a Cosserat segment. By defining length along the filament centre-line for the Voronoi regions, rather than through the coupling bond, we define curvature consistently with the Cosserat model, which has a significant effect when the coupling region is very short and $|\widehat{\bm{d}}_3\cdot \bm{Q}_{\mathrm{MR}}^T\hat{\bm{l}}_{\mathrm{MR}}| \ll |\hat{\bm{l}}_{\mathrm{MR}}|$. Having introduced these curvatures, we can define bending and twisting torque forces by
\begin{equation*}
    \bm{\tau}_{\mathrm{Ell}} = -\frac{\hat{\bm{B}} \hat{\bm{\kappa}}_{\mathrm{Ell}}}{\mathcal{E}_{\mathrm{Ell}}^3} + \frac{\hat{\bm{\kappa}}_{\mathrm{Ell}} \cross\hat{\bm{B}}\hat{\bm{\kappa}}_{\mathrm{Ell}}}{2\mathcal{E}_{\mathrm{Ell}}^3}\hat{D}_{\mathrm{Ell}} \, ,
\qquad
\mbox{and}
\qquad
\bm{\tau}_{\mathrm{Cos}} = \frac{\hat{\bm{B}} \hat{\bm{\kappa}}_{\mathrm{Cos}}}{\mathcal{E}_{\mathrm{Cos}}^3} + \frac{\hat{\bm{\kappa}}_{\mathrm{Cos}} \cross \hat{\bm{B}} \hat{\bm{\kappa}}_{\mathrm{Cos}}}{2\mathcal{E}_{\mathrm{Cos}}^3}\hat{D}_{\mathrm{Cos}} \, ,
\end{equation*}
where the matrix logarithm is calculated using Rodrigues' formula, in a manner identical to that used in the Cosserat model. All that is left to consider is the evolution of the orientation basis of the coupling, which we will consider to have angular velocity $\bm{\omega}_{\mathrm{MR}}$ defined with respect to $\bm{Q}_{\mathrm{MR}}$. The resulting force takes the form 
\begin{eqnarray*}
    \bm{\tau}_{\mathrm{MR}} =  \frac{\hat{\bm{B}} \hat{\bm{\kappa}}_{\mathrm{Ell}}}{\mathcal{E}_{\mathrm{Ell}}^3} 
    -\frac{\hat{\bm{B}} \hat{\bm{\kappa}}_{\mathrm{Cos}}}{\mathcal{E}_{\mathrm{Cos}}^3} + \frac{\hat{\bm{\kappa}}_{\mathrm{Ell}} \cross\hat{\bm{B}}\hat{\bm{\kappa}}_{\mathrm{Ell}}}{2\mathcal{E}_{\mathrm{Ell}}^3}D_{\mathrm{Ell}}
    + \frac{\hat{\bm{\kappa}}_{\mathrm{Cos}} \cross \hat{\bm{B}} \hat{\bm{\kappa}}_{\mathrm{Cos}}}{2\mathcal{E}_{\mathrm{Cos}}^3}D_{\mathrm{Cos}} \\ \nonumber + 
    \left(\bm{Q}^T_{\mathrm{MR}} \bm{t}_{\mathrm{MR}} \cross \hat{\bm{S}}\bm{\sigma}_{\mathrm{MR}}\right) |\hat{\bm{l}}_{\mathrm{MR}}| + \frac{ \hat{\bm{I}}\bm{\omega}_{\mathrm{MR}}}{e_{\mathrm{MR}}^2}\pdv{e_{\mathrm{MR}}}{t} + \left(\frac{ \hat{\bm{I}}\bm{\omega}_{\mathrm{MR}}}{e_{\mathrm{MR}}}\right) \cross \bm{\omega}_{\mathrm{MR}} \, .
\end{eqnarray*}
By proceeding in this manner we are able to fully capture the shear/stretch and bend/twist of the coupling region in a way that is consistent with both of the models we are coupling.

\section{Example cases}
\label{secnumexamples}

\begin{figure*}[t]
  \centering
  \includegraphics[width=0.9\textwidth]{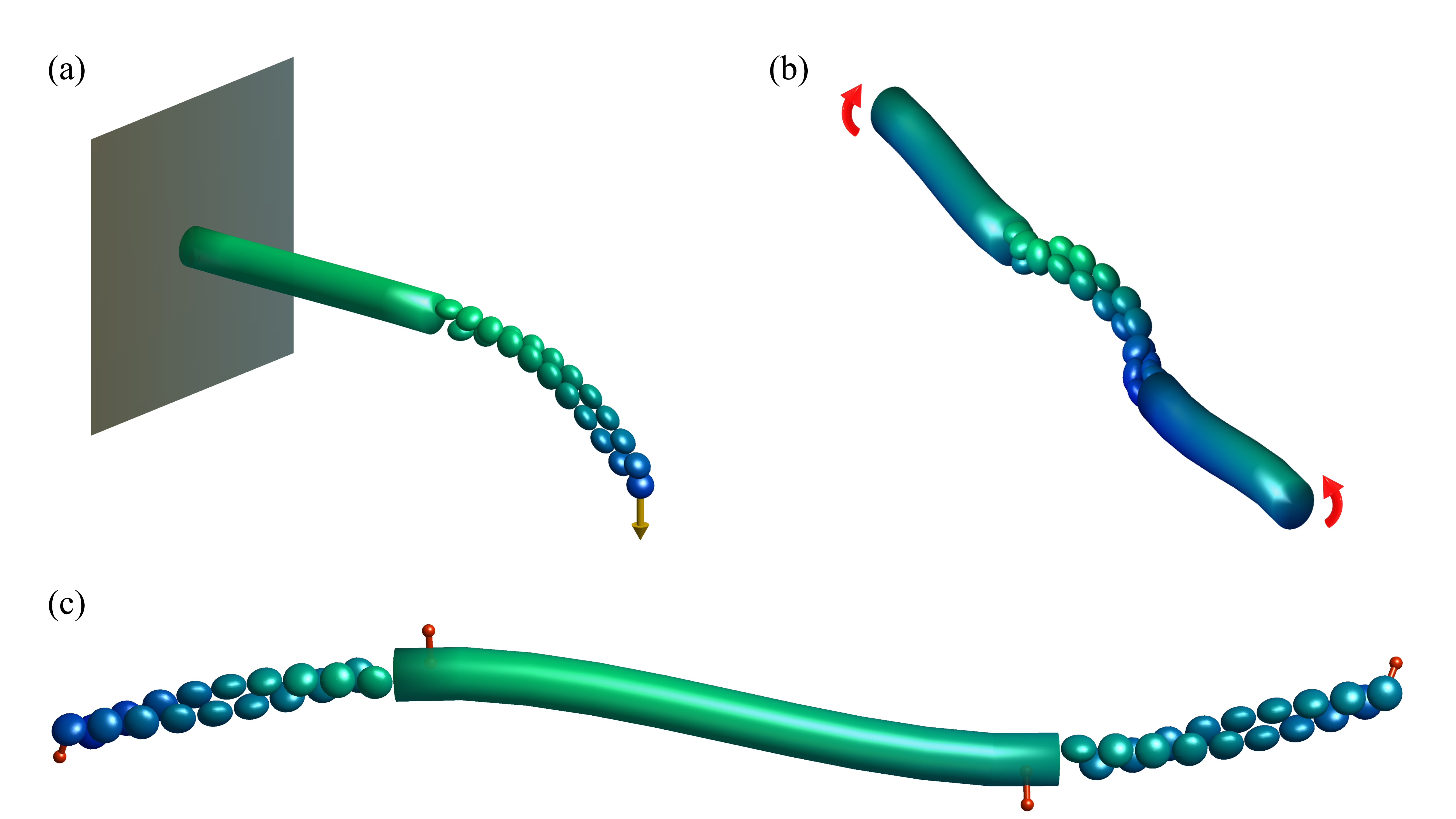}
  \caption[Illustration of example cases for multi-resolution actin filaments.]{ {\it Illustration of example cases for multi-resolution actin filaments.} \hfill\break
  (a) {\it A cantilever filament attached to a wall (grey) at one end with a downward force at the other, causing it to bend.} \hfill\break 
  (b) {\it A filament under compression and twisting forces, causing a helical buckling.} \hfill\break (c) {\it A filament attached to anchor points (red dots) by external linker bonds.}}
  \label{figure3}
\end{figure*}

Having established how to consistently couple a Cosserat segment to the ellipsoid monomers, we now study whether the additional degrees of freedom and coupling forces adversely affect the dynamic properties of filaments. In Figure~\ref{figure3}, we present some of our illustrative example cases: (a) a filament attached to a wall at one end under a cantilever force, (b) one under compression and twisting to cause helical buckling, and (c) one acted on by external linker forces. For each case we study the dynamics and static equilibrium (where dissipation is employed) of F-actin filaments resolved in varying degrees of detail.

In the preceding paper~\cite{Floyd:2022:SBSI}, the experimentally-based model parameters for the Cosserat modelling of shear, stretch, bending, and twisting rigidities are given. F-actin is described as a filament with cylindrical cross-section for comparison with other variational models. However, De La Cruz et al.~\cite{DeLaCruz2010OriginFilaments} infer the parameters for the Ellipsoid model by matching them against the fluctuations of an all-atom simulation of F-actin. While this `bottom up' approach can be expected to give an accurate picture of the relative inter-monomer forces, the macroscopic behaviour does not exactly match up with experimental measures \cite{Schramm2017ActinDissociationb,Schramm2019PlasticFilamentsb}, with lower filament rigidity observed under bending and twisting. 

To address this issue, we scale the parameters for the models such that the bending rigidity for the Cosserat and Ellipsoid models are matched. For cantilever filaments, if a load force $F$ causes a downward deflection $\Delta L_{\mathrm{D}}$, then the bending rigidity is given by
\begin{equation*}
   B_{1} = B_{2} = \frac{F L^3}{3 \, \Delta L_\mathrm{D}} \, ,
\end{equation*}
where $L$ is the length of the filament~\cite{gere1997mechanics}. These rigidities $B_{1},$ $B_{2}$ are the first and second diagonal elements of the bending rigidity matrix $\bm{B}$ for the Cosserat model. In addition, the twisting rigidities are matched using the relation
\begin{equation}
    B_{3} = \frac{\tau L}{\Delta \theta} \, ,
\label{B3formula} 
\end{equation}
where $\Delta \theta$ is the twist angle under a torsional force $\tau$. We also set the extension/compression rigidity by extending an ellipsoidal filament, fixing one end and applying a force on the other, and using the formula
\begin{equation}
    S_{3} = \frac{F L}{\Delta L}
\label{S3formula}
\end{equation}
where $\Delta L$ is the length the filament extends under force $F$. This rigidity $S_{3}$ is the last diagonal element of the bending rigidity matrix $\bm{S}$ for the Cosserat model. Accurate direct measurement of $S_{1}$ and $S_{2}$ from the Ellipsoid model is difficult, so we use an approximation based on the shearing behaviour of a uniform filament. The specifics of the verification simulations and these shearing parameter choices, can be found in Appendix~\ref{appendixA}, along with a discussion of how the masses and inertia of the models are coupled. While the masses and inertia will not affect the final equilibrium configurations reached in our example cases, they must still be matched to ensure consistent dynamic behaviour along the filament length. 

\subsection{Cantilever filament}\label{seccantilever}

We begin with a simple cantilever filament case, where F-actin is attached to a wall at one end, as shown in Figure~\ref{figure3}(a). This is achieved by setting both the angular and linear velocity at the wall to zero throughout the simulation. At the other end, a downward (z-direction) force is applied, causing the filament to bend. This allows us to establish the behaviour of the multi-resolution model under a simple bending deformation, and also resembles situations where actin filaments are attached to the cell membrane. 

\begin{figure*}[t]
	\centering
	\includegraphics[width=0.9\textwidth]{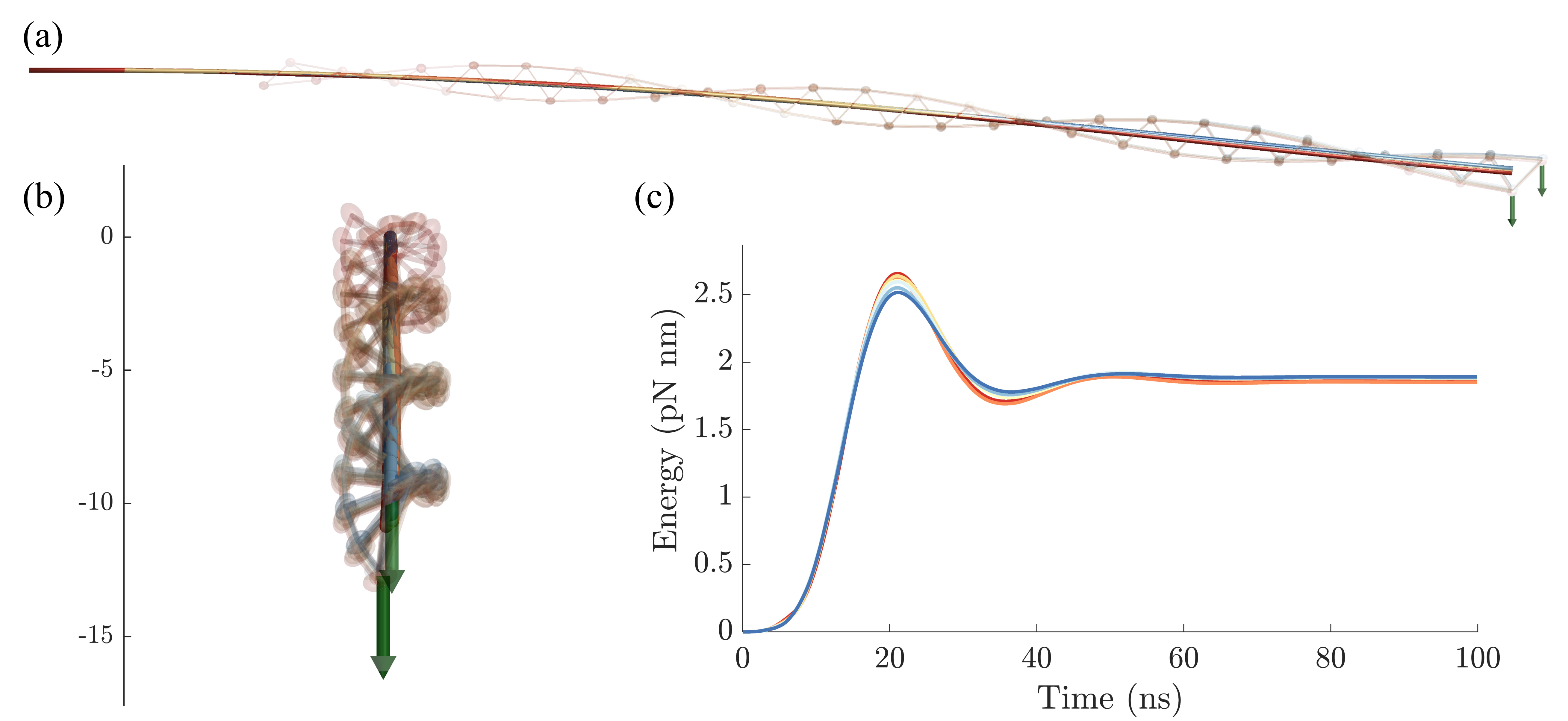}
	\caption[Multi-resolution actin filament under a cantilever force.]{(a) { \it Multi-resolution actin filament under a cantilever force. We overlay the results of $7$ simulations, with the Cosserat model used to resolve the filament for between $20$--$140\,${\rm nm}, with the remainder resolved using the Ellipsoid model, giving a $160\,${\rm nm} long filament in each case. The colours vary through red-yellow-blue, with red almost entirely monomer-based and blue mainly rod-based.} \hfil\break (b) {\it View along the filament from the right of panel} (a). \hfill\break (c) {\it The total filament energy $($in {\rm pN$\,$nm}$)$ varying with time $($in {\rm ns}$)$ as the system approaches the equilibrium.}}
\label{figure4}
\end{figure*}

In Figure~\ref{figure4}, we see filaments of length $L = 160\,$nm under a cantilever force of magnitude $F = 2\,$pN, reaching an equilibrium position with the filament bent downwards. The final spatial coordinates are visualised in Figure~\ref{figure4}(a), with the results from 7 simulations of varying resolution overlayed. For the Cosserat model we do not visualize full cylinders, but only their centre-line. For the Ellipsoid model we plot the monomer centres of mass, connected by thin cylinders representing lateral and longitudinal interactions. We also fit a curve to approximate the ellipsoidal filament centre-line to allow easier comparison between resolutions. We see that the results from these simulations overlap, indicating the accuracy of the multi-resolution methodology. This is confirmed in Figure~\ref{figure4}(b) where the same results are visualised from a different viewpoint. Finally, Figure~\ref{figure4}(c) illustrates that the dynamical properties are also matched well for all cases. For all levels of resolution, we have agreement with the theoretically expected deflection
\begin{equation*}
    \Delta L_\mathrm{D} = \frac{F L^3}{3 E I} \approx 10\, {\rm nm}
\end{equation*}
where $E=2000\,$pN$\,$nm$^{-2}$ is the Young's modulus, and $I = A^2 / 4 \pi \approx 120\,$nm$^4$ is the area moment of inertia, which corresponds to the first and second diagonal elements of the bending rigidity matrix $\bm{B}$ of the Cosserat model. This confirms that the bending parameters for the two models are correctly coupled. 

\subsection{External Linker Forces}
\label{secexternallinkers}

Having investigated the effects of a force on a filament anchored to a wall, we now consider a free filament attached to multiple external linkers, causing non trivial bending behaviour, as shown in Figure~\ref{figure3}(c). This is a more realistic study of the forces an F-actin filament would experience in the cytoskeleton. We define a general binding site in terms of the existing Cosserat segment endpoints and orientation basis as follows
\begin{equation*}
	\bm{r}_{\mathrm{ext}} = \bm{r}_{i} + A_0 (\bm{r}_{i+1} - \bm{r}_{i}) + R (\bm{d}_{1,i} \cos(\phi_0) + \bm{d}_{2,i} \sin(\phi_0)) 
\end{equation*} 
where $A_0 \in [0, 1]$ defines the cross-sectional location of the binding site along the length of the segment, $R$ is its radial distance from the filament centre-line, and $\phi_0$ is the angle in the cross-section the point forms relative to $\bm{d}_{1,i}$. This allows torsional effects of linkers on the Cosserat filament to be accurately modelled. As actin has high shear rigidity it is reasonably safe to assume that $\bm{d}_{3,i}$ is almost parallel to the segment tangent, meaning this rigid approximation to the position of the binding site will hold even for long Cosserat segments. If we have a force $\bm{F}_{\mathrm{ext}}$ at $\bm{r}_{\mathrm{ext}}$ due to the external linkers, this can be translated to the existing filament coordinates by considering how it acts on the midpoint of the Cosserat segment
\begin{equation*}
    \bm{F}_{i} = \bm{F}_{i+1} = \frac{1}{2} \bm{F}_{\mathrm{ext}} \, , \quad 
    \bm{C}_{\mathcal{L}}^{i} = \bm{Q}^T_{i} \left(\left( \bm{r}_{\mathrm{ext}} - \frac{1}{2}(\bm{r}_{i+1} + \bm{r}_{i})\right) \cross \bm{F}_{\mathrm{ext}}\right)
\end{equation*}
with the external torque $\bm{C}_{\mathcal{L}}^{i}$ acting on the orientation basis $\bm{Q}_{i}$, enabling twisting of the filament to take place.
\begin{figure*}[t]
	\centering
	\includegraphics[width=0.9\textwidth]{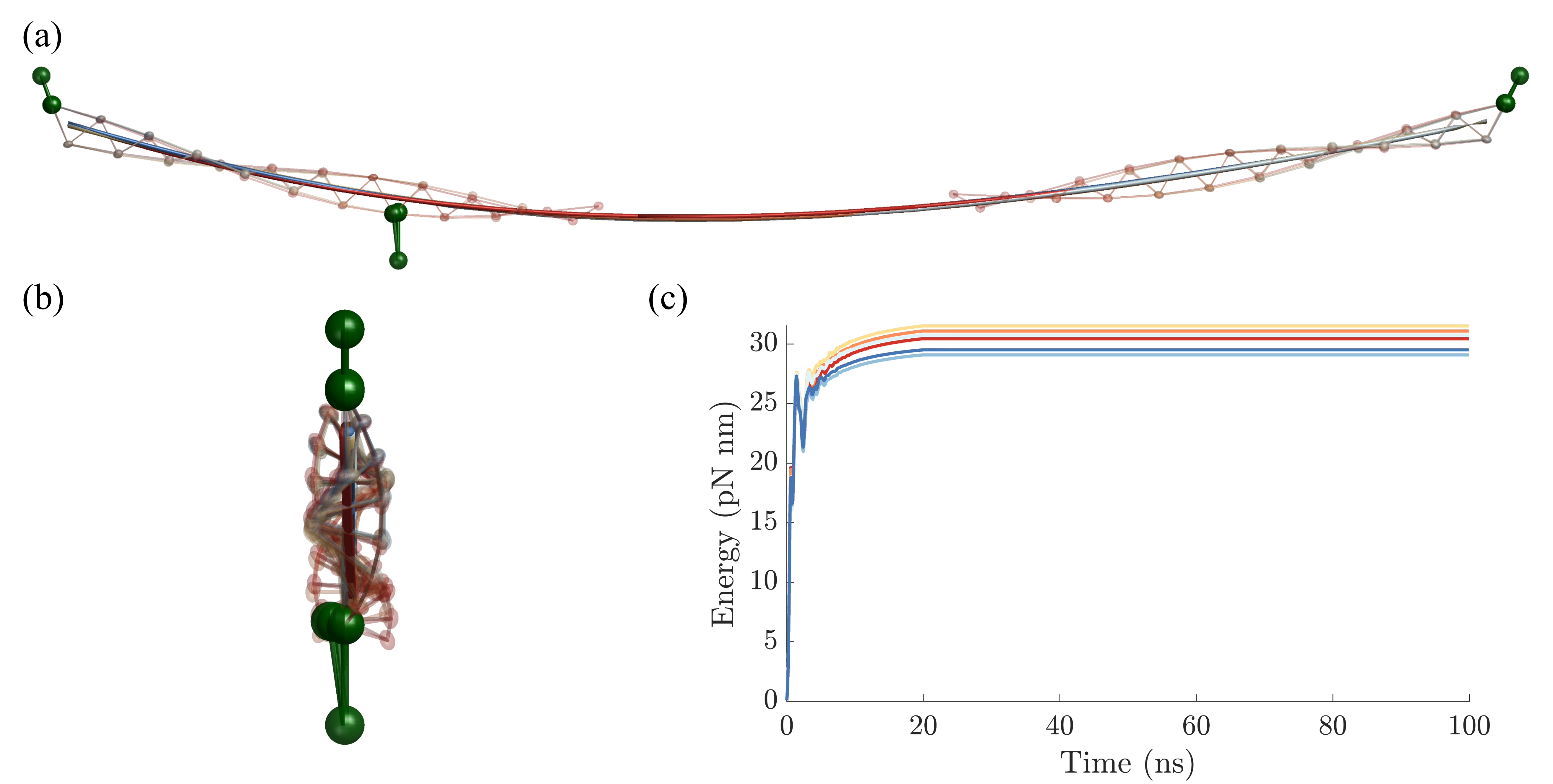}
	\caption[Multi-resolution actin filament attached to $3$ planar external linkers.]{(a) { \it Multi-resolution actin filament attached to $3$ planar external linkers. We overlay the results of $6$ simulations, with the Cosserat model used to resolve the filament for between $30$--$130\,${\rm nm} in the middle section, with the remainder resolved using the Ellipsoid model at both ends of the filament. The colours vary through red-yellow-blue, with red almost entirely monomer-based and blue mainly rod-based. } \hfill\break (b) {\it View along the filament from the right of panel} (a). \hfill\break (c) {\it The total filament energy $($in {\rm pN$\,$nm}$)$ varying with time $($in {\rm ns}$)$ as the system approaches the equilibrium.}}
	\label{figure5}
\end{figure*}
In Figure~\ref{figure5}, we see F-actin filaments of length $160$~nm with 3~linkers attached from fixed external points. For this case, we place the 3 linkers in the same plane, illustrating a simple bending situation with low filament torsion. We have excellent structural agreement between the models, despite the binding location switching from the Cosserat to the Ellipsoid model as the resolution is varied. 
\begin{figure*}[t]
	\centering
	\includegraphics[width=0.9\textwidth]{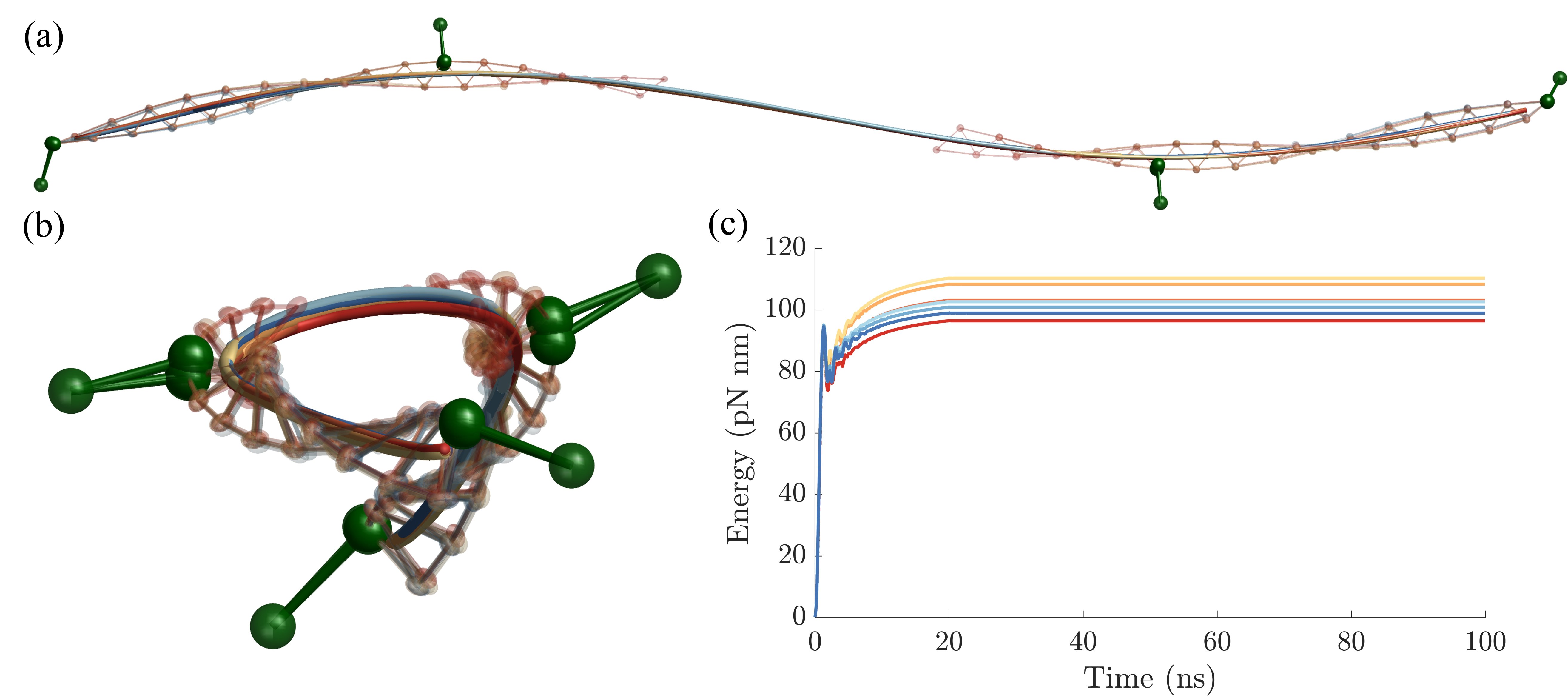}
	\caption[Multi-resolution actin filament attached to $4$ non-planar external linkers.]{(a) { \it  Multi-resolution actin filament attached to $4$ non-planar external linkers. We overlay the results of $8$ simulations, with the Cosserat model used to resolve the filament for between $30$--$170\,${\rm nm} in the middle, with the remainder resolved using the Ellipsoid model at both ends. The colours vary through red-yellow-blue, with red almost entirely monomer-based and blue mainly rod-based.}
	\hfill\break (b) {\it View along the filament from the right of panel} (a). \hfill\break (c) {\it The total filament energy $($in {\rm pN$\,$nm}$)$ varying with time $($in {\rm ns}$)$ as the system approaches equilibrium.}}
	\label{figure6}
\end{figure*}
We extend this to a more complex example in Figure~\ref{figure6}, where an F-actin filament of length $210$~nm has 4 linkers attached to its from fixed points. In this case, rather than all of these linkers lying in a single plane, we place them more randomly in space, inducing larger torques and filament deformation. While agreement across model with differing resolutions is still good, deviations between the equilibrium configurations are more noticeable, which is also contributes to the larger variation in equilibrium filament energy in Figure~\ref{figure6}(c).

\subsection{Twisted Filament}\label{sectwist}

While the multi-resolution model performs well under linear forces which cause it to bend, there is not perfect agreement, which is partly due to the double helical structure of actin. When the actin double helix is wound more tightly this is known as over-twisting, while when it is unravelled in the opposite direction it is under-twisted. It is easier to under-twist than over-twist the filament. While we matched the parameters for the two models by directly measuring rigidities from the Cosserat model, it is difficult to incorporate the helicity of F-actin into the Cosserat model without introducing complex extra energy terms~\cite{Floyd:2022:SBSI}. A good way of highlighting the difference in the models is to analyse a filament under a simple twist. We clamp both ends of $\sim 180\,$nm long filaments and attach a linker to the middle, away from the centre-line so that a torque is created. 
\begin{figure*}[t]
\hskip 0.5mm
\includegraphics[width=0.45\linewidth]{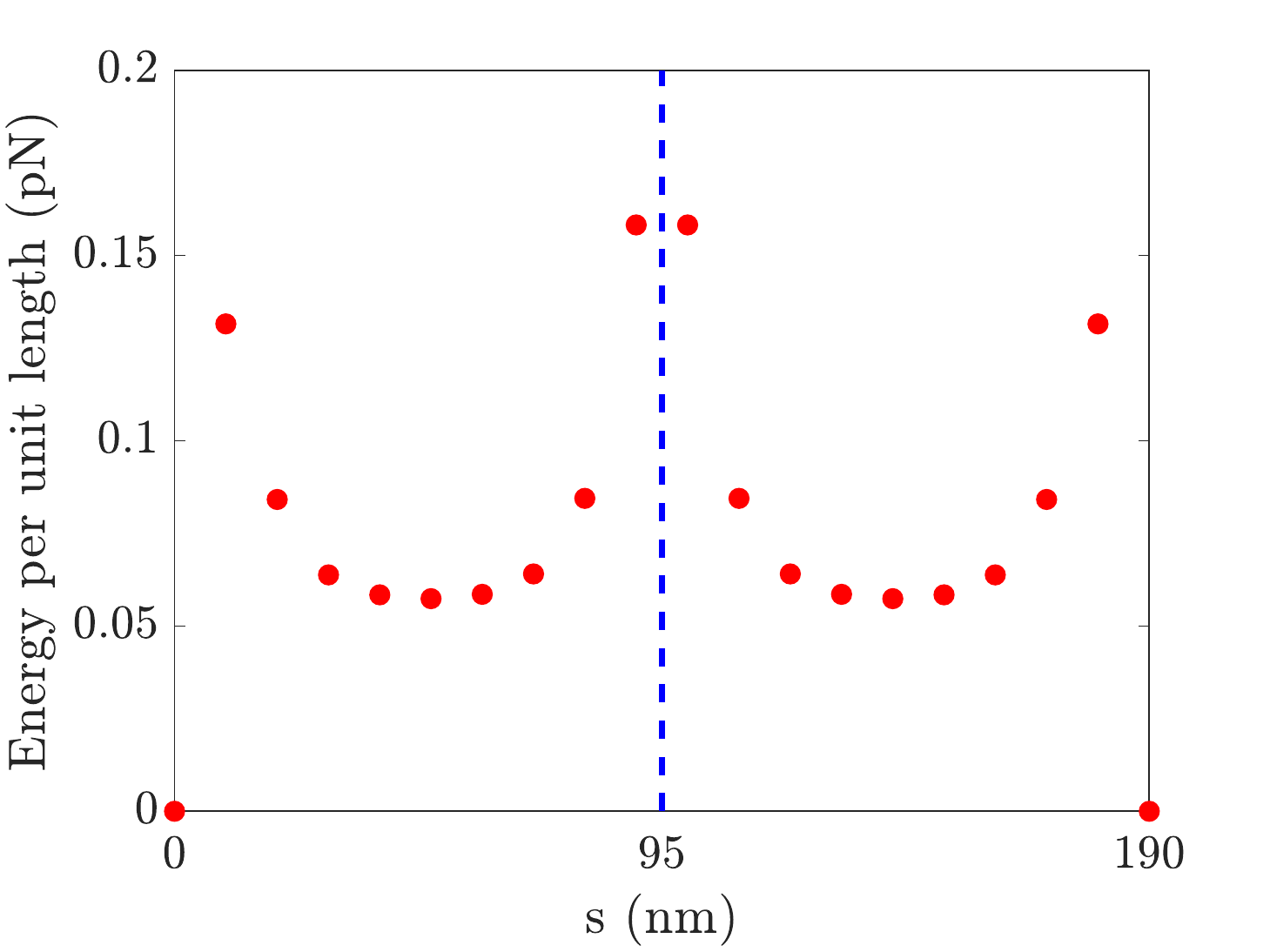}
\hskip 8mm
\includegraphics[width=0.45\linewidth]{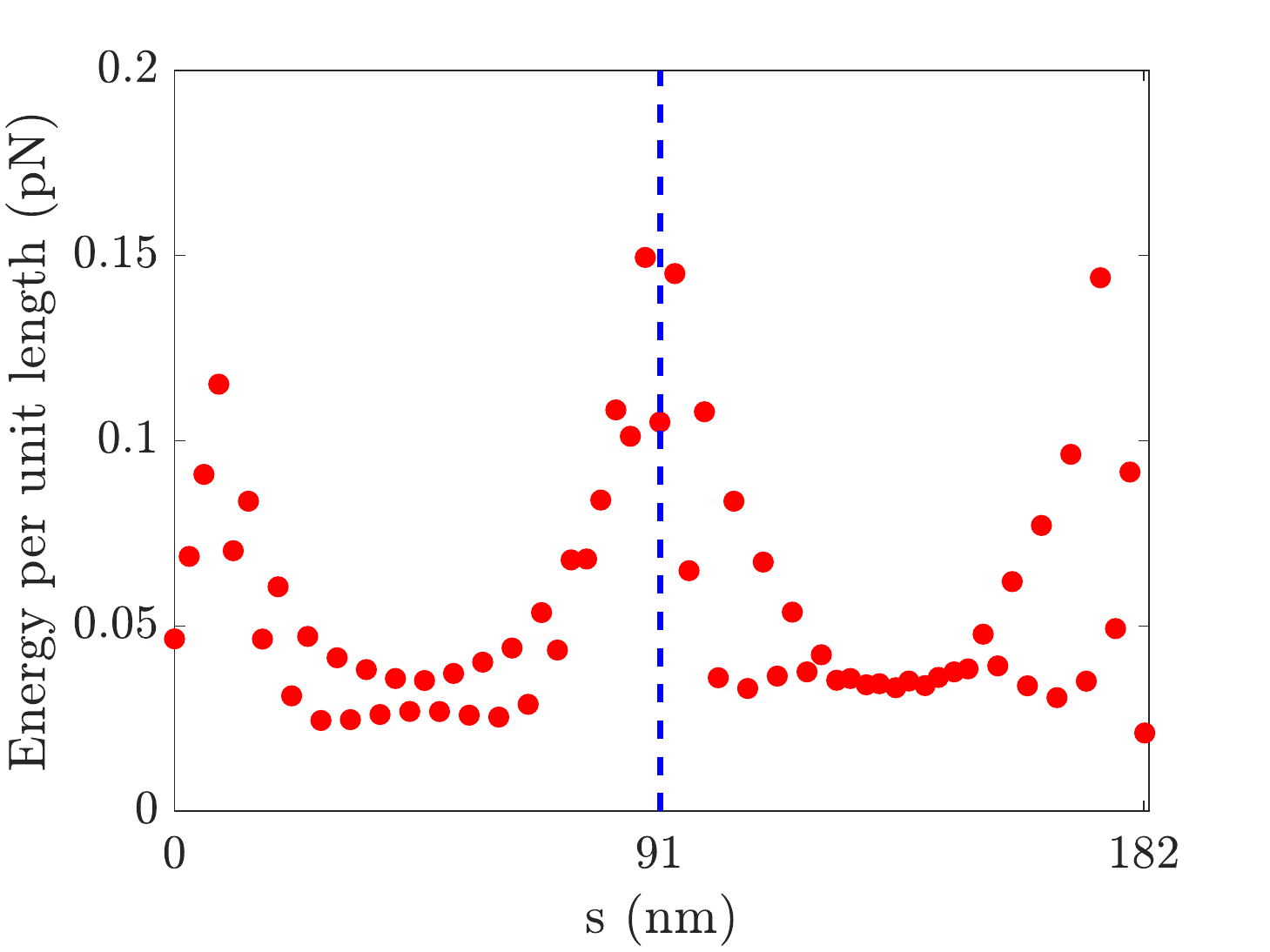}
\vskip -5.8cm
(a) \hskip 7.8cm (b) \hfill\break
\vskip 4.8cm
\caption[The spatial distribution of potential energy under torsion for Cosserat and ellipsoidal filaments]{ {\it The spatial distribution of potential energy under torsion for:} (a) {\it the Cosserat model with $19$ segments; and}\/ (b) {\it the Ellipsoid model with $67$ ellipsoids. In both cases we clamped the ends of a filament and attach an external linker to the middle $($blue dashed line$)$. In the ellipsoid case the linker is attached to a monomer centre, while in the Cosserat model it is attached with the same radial distance from the centre-line, inducing an equivalent torque. The red markers indicate the energy per unit length.}}
  \label{figure7}
\end{figure*}
In Figure~\ref{figure7}, we see the comparison between these filaments, with one resolved using only the Cosserat model, and the other entirely the Ellipsoid model. An external linker with zero length and spring constant $100\,$pN$\,$nm$^{-1}$ is attached to the middle of both filaments to induce a torque and slight bend, with the system allowed to reach equilibrium. We see that the spatial distribution of energy for the Cosserat model is symmetric on either side of location the central force is applied. However, there is a difference on either side of the binding site for the Ellipsoid model. This is due to the fact that one side is over-twisted (right side in second panel) with roughly $10\%$ higher total energy than the other side which is under-twisted (left side in second panel). In addition to variable twisting rigidity, the Ellipsoid model also exhibits twist-stretch coupling behaviour; when the filament is over-twisted it contracts more than when under-twisted. 

To illustrate this more clearly, we consider an example case shown in Figure~\ref{figure3}(b), where we hold the ends of the filament and move them together to compress the filament, causing it to buckle once the force becomes too great. At the same time, the filament ends are twisted in opposite directions along its length, resulting in helical buckling. As well as being a good setting to establish the torsional behaviour of the multi-resolution model, this has been observed in actin inside filopodia~\cite{Leijnse2015HelicalTraction}.   
\begin{figure*}[t]
\centering
\includegraphics[width=0.92\textwidth]{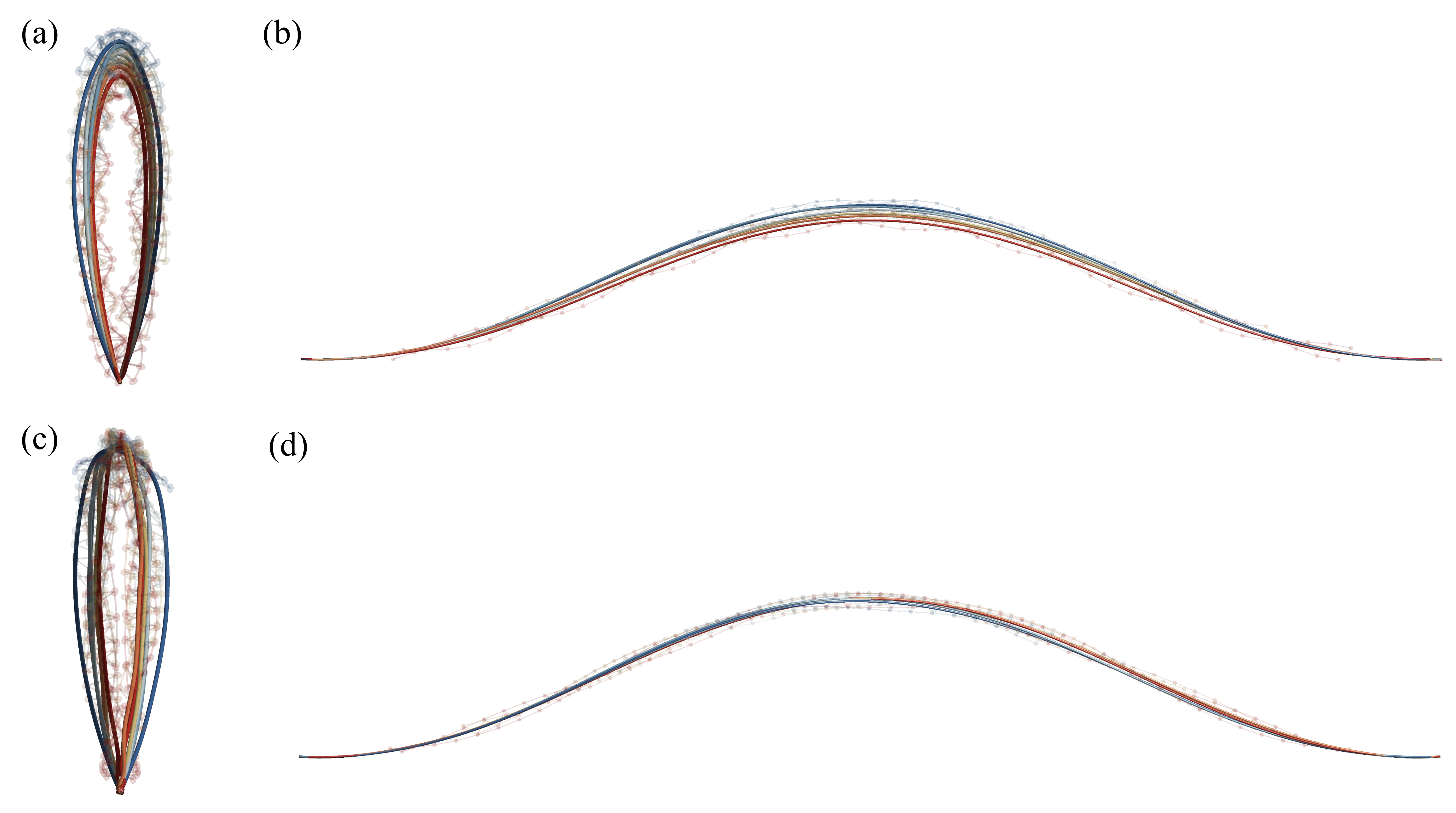}
\caption[Multi-resolution actin filaments under compression and twist.]{{\it Multi-resolution actin filaments under compression and twist. Each $300\,$nm long filament is twisted $180^o$ at each end resulting in a full rotation, and compressed by $7.5$nm at each end, giving a $5\%$ reduction in length which causes buckling. We overlay the results of $7$ simulations, with the Cosserat model used to resolve the filament for between $40$--$240\,\mathrm{nm}$ split evenly between both ends, with the remainder in the centre resolved using the Ellipsoid model. The colours vary through red-yellow-blue, with red almost entirely monomer-based and blue mainly rod-based.\hfill\break
{\rm(a) and (b)} Over-twisted viewed from end and side, respectively. \hfill\break 
{\rm(c) and (d)} Under-twisted viewed from end and side, respectively.}}
\label{figure8}
\end{figure*}

In Figure~\ref{figure8}, we see a $300\,$nm filament under compression and torsion. In Figure~\ref{figure8}(a) and~\ref{figure8}(b), we over-twist the filament, meaning the double helical ellipsoidal section of F-actin becomes more tight. We see that the vertical height of the buckled filament is larger in the lower resolution filaments (results visualised in blue). In Figure~\ref{figure8}(c) and~\ref{figure8}(d), the filaments are set up in an identical manner to the previous case, but with the direction of twist reversed, under-twisting the filament. We again see the lower resolution filaments displaying larger deformations. To account for these twisting anisotropies in the Ellipsoid model, we make two amendments to the standard Cosserat model: (i) having curvature dependent rigidities, and (ii) incorporating a twist-stretch coupling.

\begin{figure*}[t]
\hskip 0.5mm
\includegraphics[width=0.49\linewidth]{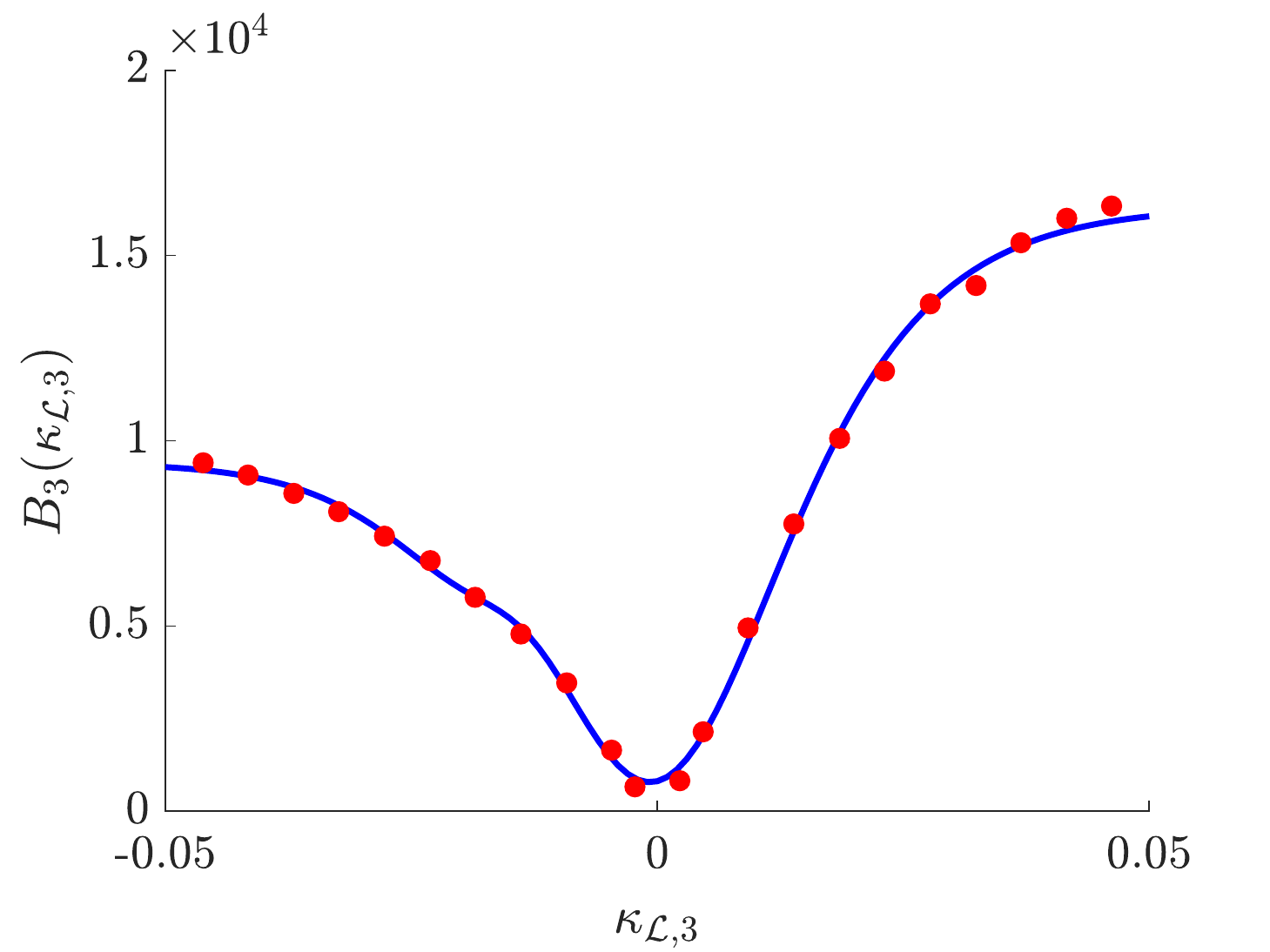}
\hskip 1mm
\includegraphics[width=0.49\linewidth]{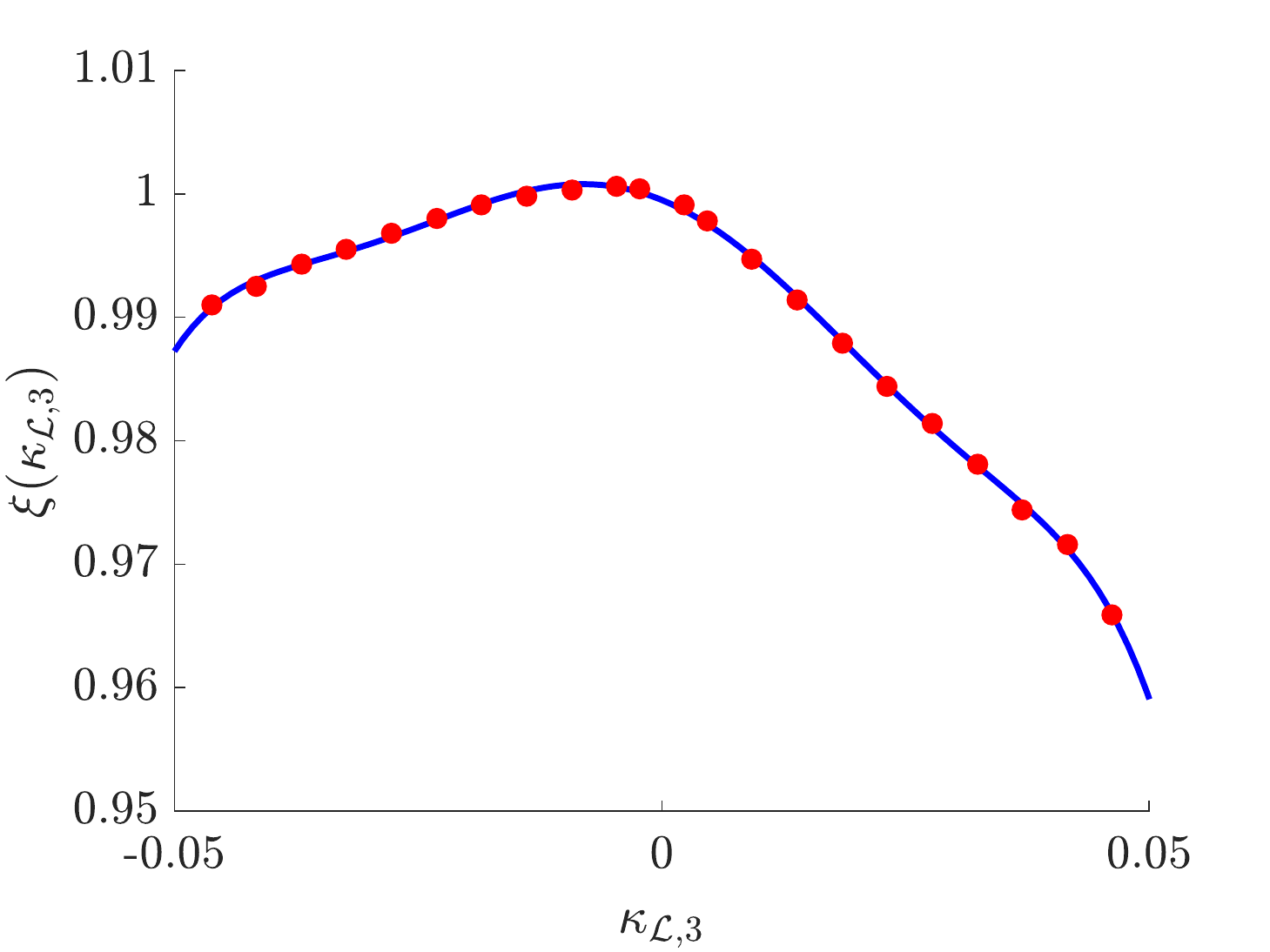}
\vskip -6.2cm
(a) \hskip 7.5cm (b) \hfill\break
\vskip 5.2cm
\centering
\caption[Twisting rigidity and stretching of ellipsoidal actin filament for varying curvature.]{{\it {\rm(a)} Twisting rigidity of ellipsoidal actin filament for varying curvature. We over-twist and under-twist a straight filament of length $\sim 75\,\mathrm{nm}$ and measure the twisting rigidity for these points (circular markers). A continuous function for $B_3(\kappa_{\mathcal{L},3})$ fitted to this data.\hfill\break
{\rm(b)} Stretching under varying curvature. We measure the extension and compression of this filament under same conditions, with continuous function  $\xi(\kappa_{\mathcal{L},3})$ fitted to data.}}
\label{figure9}
\end{figure*}

We let the bending matrix $\bm{B}$ for the Cosserat model vary with in curvature $\hat{\bm{\kappa}}_{\mathcal{L}}$. For simplicity, we limit ourselves to the twisting curvature $\hat{\kappa}_{\mathcal{L},3}$ and rigidity $B_3$ in this example. We formulate a continuous function
\begin{equation}
\begin{split}
    B_3(\hat{\kappa}_{\mathcal{L},3}) = & \left(\frac{B_{+} + B_{-}}{2}\right) + \left(\frac{B_{+} - B_{-}}{2}\right) \tanh\left(\frac{\hat{\kappa}_{\mathcal{L},3} + \lambda_1}{\lambda_2}\right)\\
    & + \lambda_3 \left(\frac{B_{+} + B_{-}}{2}\right) \left( \tanh\left(\frac{\hat{\kappa}_{\mathcal{L},3} - \lambda_4}{\lambda_5}\right) + \tanh\left(\frac{-\hat{\kappa}_{\mathcal{L},3} - \lambda_6}{\lambda_7}\right)\right),
\end{split}
\label{formulaB3}
\end{equation}
where $\lambda_i$, $i=1, 2, \dots, 7$ are a set of parameters to be inferred. The first line of the equation tends to $B_{-}$ and $B_{+}$ when $\hat{\kappa}_{\mathcal{L},3} \rightarrow -\infty$ and $\hat{\kappa}_{\mathcal{L},3} \rightarrow \infty$ respectively, while the second line of the equation tends to $0$ when $\hat{\kappa}_{\mathcal{L},3} \rightarrow \pm \infty$. This function choice is motivated by direct measurements of $B_3(\hat{\kappa}_{\mathcal{L},3})$ from the Ellipsoid model for a range of fixed twisting angles. We present data in Figure~\ref{figure9}(a) together with the continuous expression $B_3(\hat{\kappa}_{\mathcal{L},3})$ for twisting rigidity which can be used in the Cosserat model.
The inferred parameters are given in Appendix~\ref{appendixB}.

Having addressed the variable twisting rigidity, we turn our attention to the relation between twisting and stretching in the Ellipsoid model. This is done by making the segment extension $e_i$ in the Cosserat model dependent on filament curvature. We define a polynomial that varies in the twist curvature
\begin{equation}
\xi(\hat{\kappa}_{\mathcal{L},3}) = \sum\limits_{i=0}^6 p_i \hat{\kappa}_{\mathcal{L},3}^i
\label{polxikappa}
\end{equation}
where $p_i$, $i = 1,2,\dots, 7$ are parameters to be inferred. This is the scale factor for twist induced filament stretching, e.g. if for a filament of length $L$ a twist of $\hat{\kappa}_{\mathcal{L},3}$ causes an extension of $\Delta L$, the scaling would be
\begin{equation*}
    \xi(\hat{\kappa}_{\mathcal{L},3}) = \frac{L + \Delta L}{L}.
\end{equation*}
We again use data collected from simulations of the Ellipsoid model, where a filament is twisted by a fixed angle, to parametrise our scaling function. We fix the spatial position of one end, and allow the other to move exclusively in the direction of the filament centre-line with fixed orientation. We measure this change in natural length under torsion, with data and fitted version of $\xi(\hat{\kappa}_{\mathcal{L},3})$ seen in Figure~\ref{figure9}(b). We use this to scale the discrete extension quantities for the Cosserat model $\mathcal{E}_i$ and $e_i$, resulting in filament compression or extension when twisted. More details for both of these extensions to the Cosserat model (including full data tables) are provided in Appendix~\ref{appendixB}.

\begin{figure*}[ht]
\centering
\includegraphics[width=0.86\textwidth]{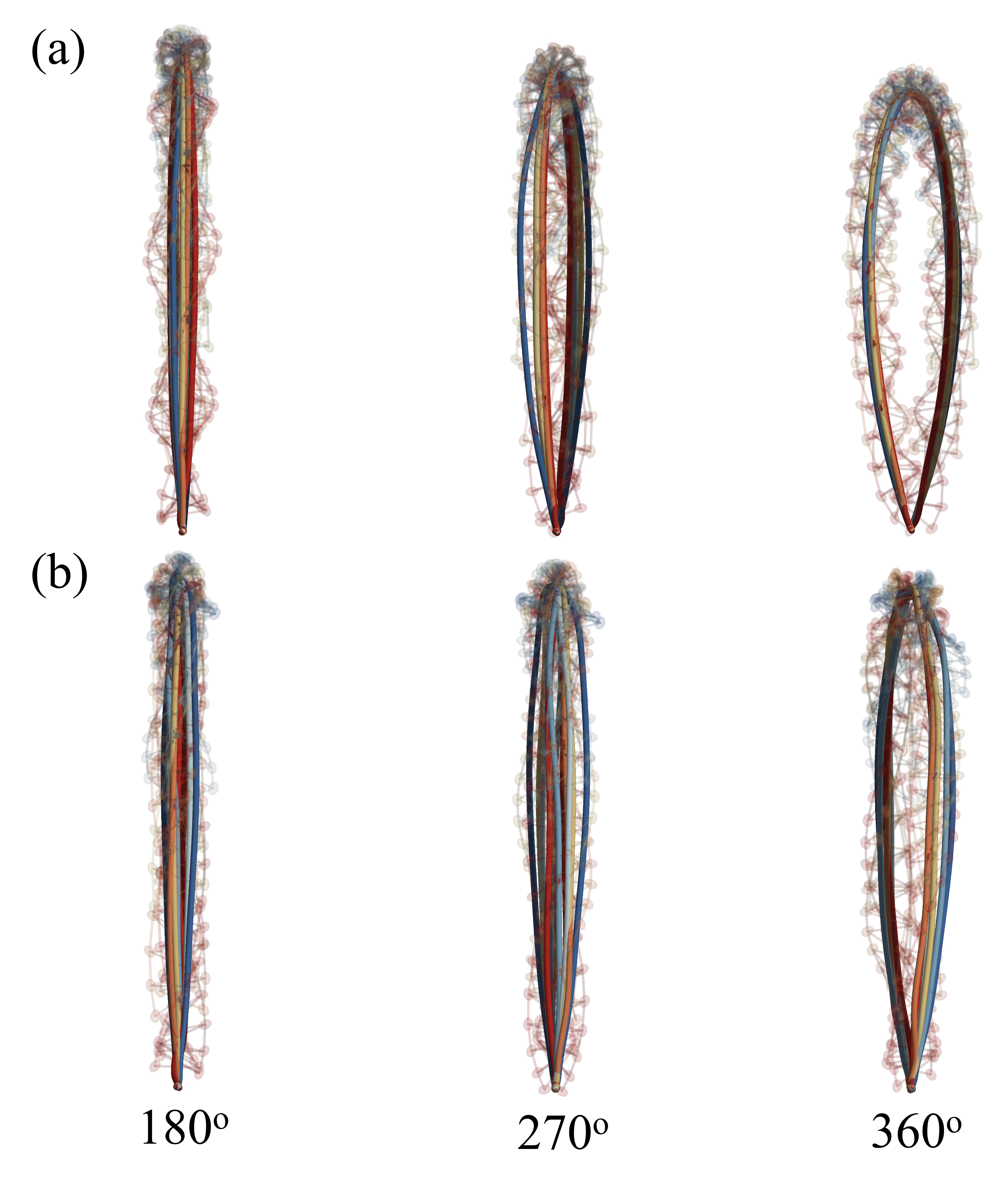}
\caption[Multi-resolution actin filaments under compression and twist, with variable twisting rigidity $B_3(\kappa_{\mathcal{L},3})$ and length scale factor $\xi(\kappa_{\mathcal{L},3})$ .]{{\it Multi-resolution actin filaments under compression and twist, with variable twisting rigidity $B_3(\kappa_{\mathcal{L},3})$ and length scale factor $\xi(\kappa_{\mathcal{L},3})$. These simulations are initialised in an otherwise identical manner to Figure~$\ref{figure8}$.\hfill\break {\rm(a)} over-twisted filaments subjected to $180^\circ$, $270^\circ$, and $360^\circ$ twists.\hfill\break{\rm(b)} under-twisted filaments subjected to $180^\circ$, $270^\circ$, and $360^\circ$ twists.}}
\label{figure10}
\end{figure*}

In Figure~\ref{figure10}, we see that the extended Cosserat model improves correspondence for the over-twist case across resolutions and over a range of different twist curvatures. However, for the under-twist case, while we observe slight improvement, at larger twists there is still lower agreement between models of differing resolution. 

\section{Discussion}
\label{secdiscussion}

We have presented a multi-resolution model for actin filaments, using a rod-based Cosserat model~\cite{Gazzola2018ForwardFilaments} for low resolution regions and a monomer-based Ellipsoid model~\cite{DeLaCruz2010OriginFilaments,Schramm2017ActinDissociationb} for high resolution regions. The coupling region is defined such that two monomers at the end of the ellipsoidal filament are attached to the final segment of a Cosserat filament. The forces at this interface constrain linear and angular movement of the models in a manner consistent with the Cosserat formulation. The parameters for the standard Cosserat model are inferred directly from simulations of the Ellipsoid model. This allows us to match both the dynamics and equilibrium structures for filaments of differing resolutions, along with having closely matching energies. 

We demonstrate this in three main types of example cases: cantilever bending, external linker forces, and twisted filaments. The multi-resolution model, with the standard Cosserat model used for low resolution regions, performs well under simple bending deformations. Agreement is notably good when these are in-plane, though for out-of-plane deformations we observe slight differences in the final equilibrium energy across filaments of varying resolutions. However, when placing the filaments under torsion, the standard Cosserat model is unable to capture the anisotropies of the Ellipsoid model, as seen in Figure~\ref{figure8}. To address this, we present an augmented Cosserat model where the twisting rigidity $B_3(\kappa_{\mathcal{L},3})$ is dependent on the twist curvature $\kappa_{\mathcal{L},3}$. This allows us to account for the fact that it is easier to under-twist actin than it is to over-twist it. In addition, we act on the length of the filament via scaling $\xi(\kappa_{\mathcal{L},3})$, again motivated by the monomer-based representation of actin, where twist-stretch behaviour is observed. We present functions for both $B_3(\kappa_{\mathcal{L},3})$ and $\xi(\kappa_{\mathcal{L},3})$, with their parameters inferred from data collected by placing the Ellipsoid model under twists of varying magnitude. When applying these extensions to the Cosserat formulation in our multi-resolution model we are able to demonstrate improved model correspondence under torsion, even when large portions of the filament have a low resolution. 

If we employ these extensions, the original energy expression~(\ref{eq:cosserat_energy}) for the Cosserat model no longer hold~\cite{Gazzola2018ForwardFilaments}. In effect, we are letting the twisting rigidity $B_3(\kappa_{\mathcal{L},3})$ and strain $\bm{\sigma}_{\mathcal{L}}(\kappa_{\mathcal{L},3})$, via the scaling $\xi(\kappa_{\mathcal{L},3})$, vary in twist curvature, resulting in the updated energy for the augmented Cosserat model
\begin{equation*}
    E = \frac{1}{2}\int\limits_{0}^{L} \bm{\kappa}_{\mathcal{L}}^T \bm{B}(\kappa_{\mathcal{L},3})\bm{\kappa}_{\mathcal{L}}\, \mathrm{d}s + \frac{1}{2}\int\limits_{0}^{L} \bm{\sigma}_{\mathcal{L}}(\kappa_{\mathcal{L},3})^T \bm{S}\bm{\sigma}_{\mathcal{L}}(\kappa_{\mathcal{L},3})\, \mathrm{d}s .
\end{equation*}
We can still gain clear insight into the bend and twist properties of the filament via $\bm{B}(\kappa_{\mathcal{L},3})$. However, it is less simple to separate the general shear and stretch properties (determined by $\bm{S}$) from the twist-stretch coupling behaviour introduced by using the non-linear $\bm{\sigma}_{\mathcal{L}}(\kappa_{\mathcal{L},3})$ strain term. Rather than having curvature dependent strain, it is also possible to create a twist-stretch coupling by introducing an extra energy term $C_3(\kappa_{\mathcal{L},3},\sigma_{\mathcal{L},3}) \,\sigma_{\mathcal{L},3} \, \kappa_{\mathcal{L},3}$, which would allow its effect to be quantified separately via the coupling rigidity $C_3(\kappa_{\mathcal{L},3},\sigma_{\mathcal{L},3})$. In the {\it Supplementary Information} we include a basic dynamic implementation, with forces based on this energy term. However, it is a non-trivial task to infer the form of $C_3(\kappa_{\mathcal{L},3},\sigma_{\mathcal{L},3})$ to accurately match the Cosserat model to the Ellipsoid model; the inclusion of this twist-stretch coupling affects the inference of the twisting rigidity $B_3(\kappa_{\mathcal{L},3})$, and these functional forms must be parametrised together. While this is beyond the scope of this paper, a systematic study to determine $\bm{B}(\bm{\kappa}_{\mathcal{L}})$, $\bm{S}(\bm{\sigma}_{\mathcal{L}})$, and $\bm{C}(\bm{\kappa}_{\mathcal{L}},\bm{\sigma}_{\mathcal{L}})$ by direct measurement from the Ellipsoid model would enable the formulation of a class of further augmented Cosserat models with energy
\begin{equation*}
    E = \frac{1}{2}\int\limits_{0}^{L} \bm{\kappa}_{\mathcal{L}}^T \bm{B}(\bm{\kappa}_{\mathcal{L}})\bm{\kappa}_{\mathcal{L}}\, \mathrm{d}s + \frac{1}{2}\int\limits_{0}^{L} \bm{\sigma}_{\mathcal{L}}^T \bm{S}(\bm{\sigma}_{\mathcal{L}})\bm{\sigma}_{\mathcal{L}}\, \mathrm{d}s +\int\limits_{0}^{L} \bm{\kappa}_{\mathcal{L}}^T \bm{C}(\bm{\kappa}_{\mathcal{L}},\bm{\sigma}_{\mathcal{L}})\bm{\sigma}_{\mathcal{L}}\, \mathrm{d}s ,
\end{equation*}
which incorporates the curvature and strain dependence of the material properties, along with complex coupled filament behaviour. This more generalised form could better capture the anisotropies of F-actin.

Additionally, this could be extended to studies of other bio-filaments with non-trivial physical properties like filopodia~\cite{Michaels2020MechanicalMorphology}, or even more flexible filaments such as hair fibres~\cite{Bergou2008DiscreteRods,Bergou2010DiscreteThreads}. Another approach to actin filament modelling is presented by Yamaoka et al.~\cite{Yamaoka2010CouplingAxis} where, rather than varying the twist rigidity $B_3(\kappa_{\mathcal{L},3})$, the intrinsic twist curvature of actin is incorporated into the formulation of a Cosserat model by considering the mismatch between the curve passing through monomer centres of mass and the filament centre-line. Alternatively, work has been done modelling cytoskeleton filaments using inextensible rod models~\cite{maxian2022hydrodynamics} while accounting for hydrodynamics. 

In the cases where the filament experiences simple bending deformations, the Cosserat model adequately captures this behaviour, with much improved efficiency when compared to the Ellipsoid model. However, for the twist cases we have observed that even when the extended Cosserat model is used, the multi-resolution model is not as accurate at low resolutions. This suggests that the required resolution for accuracy is variable, and would benefit from the formulation of an adaptive multi-resolution model. In such an adaptive framework, a filament could be initialised entirely using the Cosserat model, with conditions determining where and when to increase the resolution of the filament to improve accuracy. These conditions could take the form of force thresholds or even a time-dependent local curvature threshold, where if a filament is in a deformed state for long enough we trigger the replacement of Cosserat segments with Ellipsoid monomers. The new ellipsoid orientations and positions can be directly calculated relative to an interpolated filament centre-line curve passing through the original Cosserat segment endpoints. This adaptive style of model has been used to increase the resolution of a bead-spring polymer~\cite{Rolls2016VaryingDynamics} when a substrate came within a threshold distance. Such models could be applied to model cofilin decoration of F-actin, with resolution increased in regions where we expect binding to take place. 

Another possible direction for future work would be to reformulate the multi-resolution model so it can be solved in a variational manner using energy minimisation, as in our preceding paper~\cite{Floyd:2022:SBSI}. While this would no longer be a dynamic model, we would be able to carry out investigations over much larger system sizes, involving many actin filaments~\cite{Floyd2019QuantifyingNetworks,Popov2016MEDYAN:Networks,Nedelec2007CollectiveFibers,freedman2017versatile}. As the coupling is defined in a manner consistent with the Cosserat model, the implementation can be extended to it without much difficulty. However, for the Ellipsoid model, the large number of harmonic bonds would adversely affect the efficiency of energy minimisation. One option would be to replace each set of harmonic interactions between a pair of monomers with a single, Cosserat style material interaction. This would allow us to maintain the monomer-based representation of actin, while essentially using just the Cosserat model energy terms to describe the entirety of the multi-resolution filament. It would be possible to infer parameters for this model directly from the Ellipsoid model, or one could also could return to atomistic MD simulation trajectories for F-actin and carry out a new fluctuation matching procedure~\cite{Schramm2017ActinDissociationb,Aydin2018MultiscaleProteins}, fitting longitudinal and lateral values for the elements of $\bm{S}$ and $\bm{B}$.  
 
\vskip 5mm

\section*{Acknowledgement}
\noindent
This work was supported by the Engineering and Physical Sciences Research Council, grant
number EP/V047469/1, awarded to Radek Erban. This work was also supported by the
National Science Foundation, grant number CHE-210268 and a Visiting Research Fellowship
from Merton College, Oxford, awarded to Garegin Papoian.

\appendix

\section{Parameterisation of the Ellipsoid and Cosserat models}
\label{appendixA}

\noindent
The Ellipsoid model was first described by De La Cruz {\it et al}\cite{DeLaCruz2010OriginFilaments}, and was later extended to cofilin decorated actin filaments\cite{Schramm2017ActinDissociationb,Schramm2019PlasticFilamentsb}. The parameters for the Ellipsoid model can be inferred from all atom MD simulations and crystal structure data\cite{Galkin2015Near-AtomicF-Actin}. The initial positions of the centres of the ellipsoid monomers are set to the centres of mass of the monomers in the F-actin chain, with their dimensions approximately matching those of actin. There are two different sets of harmonic interactions between monomers; lateral (between monomers in adjacent helical chains) and longitudinal (between monomers in the same helical chain). We set the harmonic bonds and their anchor points using the following method. If we have two adjacent ellipsoid monomers, we uniformly sample a point on the surface of the first one. This process starts by considering an ellipsoid centred at the origin, and sampling the elements of a vector $\bm{n} = [n_1,n_2,n_3]$ from a unit normal distribution, {\it i.e.} $n_k \sim N(0,1)$ for $k=1,2,3.$ With the ellipsoid principle radial lengths $a$, $b$, and $c$ (where $a>b>c$) oriented in the $x$, $y$, and $z$ directions, respectively, we define a normalised vector $\bm{q}=\bm{n}/|\bm{n}|$ and we accept point $\bm{r} = [a q_1,b q_2,c q_3]$ with probability
\begin{equation*}
p = c \, \sqrt{\left(\frac{q_1}{a}\right)^2 + \left(\frac{q_2}{b}\right)^2 + \left(\frac{q_3}{c}\right)^2}.
\end{equation*}
This acceptance-rejection step is necessary to sample a uniformly distributed point on the surface of the ellipsoid. The sampled point is then translated and rotated to the correct position on the surface of the monomer. For example, if this is the end point $\bm{r}^{k}_{i}$ of the $k^{\mathrm{th}}$ bond on the surface of the $i^{{\rm th}}$ monomer, then we have $\bm{r}^{k}_{i} = \bm{\Bar{R}}_{i} + \bm{Q}_i^T \bm{r},$ where $\bm{\Bar{R}}_{i}$ and $\bm{Q}_i$ are the centre of mass and orientation, respectively, of the $i^{{\rm th}}$ monomer. Next, we connect the point on the $i^{\mathrm{th}}$ monomer to the surface of the adjacent $j^{\mathrm{th}}$ monomer using the line parallel to the vector connecting the centres of monomers. Its equation can be written as 
\begin{equation*}
\bm{r} = \bm{r}^{k}_{i} + \lambda \, \bm{l} \, , \quad 
\mbox{where} \quad \bm{l} = \frac{\bm{\Bar{R}}_i - \bm{\Bar{R}}_j}{|\bm{\Bar{R}}_i - \bm{\Bar{R}}_j|}.
\end{equation*}
The equation for points, $\bm{r}$, on the $j^{\mathrm{th}}$ monomer surface is given by $(\bm{r} - \bm{\Bar{R}}_{j})^T \bm{Q}_j^T \bm{D}\bm{Q}_j (\bm{r} - \bm{\Bar{R}}_{j}) = 1,$ where $\bm{D} = \mathrm{diag}(1/a^2,1/b^2,1/c^2)$. Substituting $\bm{r} = \bm{r}^{k}_{i} + \lambda \, \bm{l}$, we obtain a quadratic equation for $\lambda$, which can be solved in $\lambda$ giving one or two real solutions if the line touches or intersects with the ellipsoid. In particular, the end point of the $k^{\mathrm{th}}$ bond on the surface of the $j^{{\rm th}}$ monomer is given by $\bm{r}^{k}_{j} = \bm{r}^{k}_{i} + \lambda \, \bm{l}$ with the root $\lambda$ chosen to minimise the bond length $|\bm{r}^{k}_{j} - \bm{r}^{k}_{i}|$. An additional acceptance-rejection constraint on the maximum tolerated bond length is then used to enforce sampling over a specified interface surface area. The values of bond rigidity parameters are taken from original papers on the Ellipsoid model \cite{Schramm2017ActinDissociationb,Schramm2019PlasticFilamentsb,Fan2013MolecularMechanics,Galkin2015Near-AtomicF-Actin} and are given in Table~\ref{table:1}, along with our chosen values for the maximum interface bond lengths.
\begin{table}[t]
\centering
\begin{tabular}{||c | c ||} 
 \hline
 \rowcolor[HTML]{ccddff} 
 Parameter & Value\\ 
 \hline\hline
 Actin dimensions\cite{Galkin2015Near-AtomicF-Actin}  &  $5.4 \times 5.4 \times 3.8$~nm  \\ 
 Filament period\cite{Fan2013MolecularMechanics}  & $71.2$~nm  \\
 Monomers per filament period\cite{Fan2013MolecularMechanics}  & 26  \\
 Rise per monomer\cite{Fan2013MolecularMechanics} & $5.52$~nm \\
 Effective interaction radius\cite{Fan2013MolecularMechanics} & $1.8$~nm \\
 Lateral interface rigidity\cite{Galkin2015Near-AtomicF-Actin} & $392.0$ $k_B T\,\mathrm{nm}^{-2}$  \\
 Longitudinal interface rigidity \cite{Galkin2015Near-AtomicF-Actin} & $582.4$ $k_B T$ $\mathrm{nm}^{-2}$  \\
 Lateral interface maximum bond length & $1.6$ $\mathrm{nm}$ \\
 Longitudinal interface maximum bond length & $2.2$ $\mathrm{nm}$ \\[0.5ex] 
 \hline 
\end{tabular}
\caption{{ \it Parameters of the Ellipsoid model\cite{Schramm2017ActinDissociationb}, along with maximum interface bond lengths.}}
\label{table:1}
\end{table}

In the preceding paper~\cite{Floyd:2022:SBSI}, the model parameters for the Cosserat model based on shear, stretch, bending, and twisting rigidities are derived based on experimental data, approximating actin as having a cylindrical cross-section for comparison with other variational models. The diameter $d$ of an actin filament is in the range\cite{Grazi1997WhatFilament} $5-7$~nm , giving it a cross-sectional area $A$ (assuming roughly circular) in the range 40-75 $\mathrm{nm}^2$. The second moment of inertia tensor is $\mathrm{diag}(I_1,I_2,I_3)$ given in Table~\ref{table:2} by considering circular cross-sectioned filaments. For an isotropic material, the matrix $\hat{\bm{B}}$ is diagonal and made up of the bending ($B_1$, $B_2$) and the twisting ($B_3$) rigidities, given in Table~\ref{table:2}. To estimate the shear modulus of actin, we considered the formula~\cite{Landau1986TheoryElasticity} $E = 2 G (1 + \nu)$, where $\nu$ is Poisson's ratio. For actin, $\nu$ has been estimated\cite{Tseng2002FunctionalProteins,Kojima1994DirectNanomanipulation} at around 0.4, giving us $G \approx 0.7$~GPa. The matrix $\hat{\bm{S}}$ is also diagonal, and comprises of shearing ($S_1$, $S_2$) and stretching ($S_3$) rigidities of an actin filament, given for circular cross-sections in Table~\ref{table:2}.
\begin{table}[t]
\centering
\begin{tabular}{||c c | c ||} 
 \hline
 \rowcolor[HTML]{ccddff} \multicolumn{2}{||c|}{Parameter} & Value \\ [0.5ex] 
 \hline\hline
 Filament Cross-sectional Area\cite{Grazi1997WhatFilament} & $A$ & $40$--$75$ $\mathrm{nm}^2$  \\
 Young's Modulus\cite{Kojima1994DirectNanomanipulation} & $E$ & $2000$ pN $\mathrm{nm}^{-2}$  \\
 Poisson's Ratio\cite{Kojima1994DirectNanomanipulation,Tseng2002FunctionalProteins} & $\nu$ & $\sim$ $0.4$ \\
 Shear Modulus\cite{Landau1986TheoryElasticity} & $G$ & $\sim$ $700$ pN $\mathrm{nm}^{-2}$  \\
 \hline 
 \multirow{2}{*}{Second Moment of Inertia }  & $I_{1,2}$ & $A^2 / 4\pi$ \\
 & $I_{3}$ & $A^2 / 2\pi$ \\
 \multirow{2}{*}{Shearing Rigidity\cite{Gazzola2018ForwardFilaments} }  & $S_{1,2}$ & $4 G A/3$  \\
 & $S_{3}$ & $EA$ \\
 \multirow{2}{*}{Bending Rigidity\cite{Gazzola2018ForwardFilaments} } & $B_{1,2}$ & $E I_{1,2}$ \\
 & $B_{3}$ & $ G I_{3}$ \\[0.5ex] 
 \hline
\end{tabular}
\caption{{\it Parameters for the Cosserat model for F-actin used in our preceding paper~\cite{Floyd:2022:SBSI}. }}
\label{table:2}
\end{table}

\section{Coupling of model parameters between Ellipsoid and Cosserat models}
\label{appendixB}

\noindent
Using the relations in Table~\ref{table:2}, we have that $B_{1} = E I_{1}$ where $E$ is the Young's modulus, and $I_{1} = A^2 / 4\pi$. In the interests of having some rough correspondence between our results and physical reality, the rigidities used in the Ellipsoid model are scaled up empirically. In Schramm et al.~\cite{Schramm2017ActinDissociationb}, it is stated that the simulated persistence length of the Ellipsoid model is underestimated by around 40\%, with a value of $7.0$~$\mu\mathrm{m}$ measured, rather than an expected $9.8$~$\mu\mathrm{m}$, so our scale factor will be $\sim 1.4$. This is based on experimental data~\cite{McCullough2008CofilinMechanicsb}, with measurements giving $E=330$~pN~$\mathrm{nm}^{-2}$, using the assumption that the actin cross-section is set to $A=38.5\, \mathrm{nm}^2$, and resulting in bending inertia $I_1=120\, \mathrm{nm}^4$. This scaling is further justified by other studies of F-actin~\cite{Isambert1995Flexibilityproteins,Ott1993MeasurementMicroscopy}, where the persistence length is stated to be in the range $8$--$20$~$\mu$m. The overall bending rigidity used in the multi-resolution model is $B_1$ = $3.9 \times 10^4$~pN~$\mathrm{nm}^2$, as seen in Table~\ref{table:4}. With these parameter choices, we maintain the relative microscopic monomer level behaviour of the filament, while still being able to match the observed macroscopic physical properties taken from experimental data. 

\begin{table}[t]
\centering
\begin{tabular}{|| c | c | c || c | c | c ||} 
 \hline
 \rowcolor[HTML]{ccddff}\multicolumn{3}{||c||}{Under-twist} &  \multicolumn{3}{|c||}{Over-twist}\\
 \hline\hline
 \rowcolor[HTML]{ccddff} $\phi$ & $\kappa_{\mathcal{L},3}$ & $B_3$  & $\phi$ & $\kappa_{\mathcal{L},3}$ & $B_3$\\ [0.5ex] 
 \hline
 $240^{\circ}$  & $-0.0555$ & $9.6 \times 10^3$~pN & $240^{\circ}$ & $0.0555$ & $1.2 \times 10^4$~pN\\
 $220^{\circ}$  & $-0.0508$ & $9.6 \times 10^3$~pN & $220^{\circ}$ & $0.0508$ & $1.4 \times 10^4$~pN\\
 \hline
 $200^{\circ}$  & $-0.0462$ & $9.4 \times 10^3$~pN & $200^{\circ}$  & $0.0462$ & $1.6 \times 10^4$~pN\\
 
 $180^{\circ}$  & $-0.0416$ & $9.1 \times 10^3$~pN & $180^{\circ}$  & $0.0416$ & $1.6 \times 10^4$~pN\\
 $160^{\circ}$  & $-0.0370$ & $8.6 \times 10^3$~pN & $160^{\circ}$  & $0.0370$ & $1.5 \times 10^4$~pN\\
 $140^{\circ}$  & $-0.0324$ & $8.1 \times 10^3$~pN & $140^{\circ}$  & $0.0324$ & $1.4 \times 10^4$~pN\\
 $120^{\circ}$  & $-0.0277$ & $7.4 \times 10^3$~pN & $120^{\circ}$  & $0.0277$ & $1.4 \times 10^4$~pN\\
 $100^{\circ}$  & $-0.0231$ & $6.8 \times 10^3$~pN & $100^{\circ}$  & $0.0231$ & $1.2 \times 10^4$~pN\\
 
 $80^{\circ}$  & $-0.0185$ & $5.8 \times 10^3$~pN & $80^{\circ}$  & $0.0185$ & $1.0 \times 10^4$~pN\\
 $60^{\circ}$  & $-0.0139$ & $4.8 \times 10^3$~pN & $60^{\circ}$  & $0.0139$ & $7.8 \times 10^3$~pN\\
 $40^{\circ}$  & $-0.0092$ & $3.5 \times 10^3$~pN & $40^{\circ}$  & $0.0092$ & $5.0 \times 10^3$~pN\\
 $20^{\circ}$  & $-0.0047$ & $1.6 \times 10^3$~pN & $20^{\circ}$  & $0.0047$ & $2.1 \times 10^3$~pN\\
 $10^{\circ}$  & $-0.0023$ & $6.6 \times 10^2$~pN & $10^{\circ}$  & $0.0023$ & $8.2 \times 10^2$~pN\\[0.5ex]
 \hline
\end{tabular}
\caption[Twisting rigidity of ellipsoidal actin filament at various curvatures.]{{\it Twisting rigidity of ellipsoidal actin filament at various curvatures.}}
\label{table:3}
\end{table}

Having accounted for bending behaviour, we now turn our focus to the twisting rigidities, which can be matched using the relation (\ref{B3formula}). 
This has been done by placing the filament of length $\sim 75\,$nm (i.e the actin pitch length) under a fixed twist in one of the end segments and measuring the torque force after the system relaxed. In Table~\ref{table:3} the value for the twisting rigidity is given for various levels of filament twisting, with the line above $200^{\circ}$ used to indicate the point up to which values can be considered reliable (for large twists the Ellipsoid model is either completely untwisted or beginning to coil out of plane and will not be physically realistic). With the scaling used, this results in maximum values of $B_{3} \approx 9.6 \times 10^3$~pN~$\mathrm{nm}^2$ for under-twist and $B_{3} \approx 1.6 \times 10^4$~pN~$\mathrm{nm}^2$ for over-twist. A simple compromise is to take the average of these values, giving a value of $B_{3} \approx 1.3 \times 10^4$~pN~$\mathrm{nm}^2$, which is used in the cantilever, external linker, and first twist example cases in Sections~\ref{seccantilever}, \ref{secexternallinkers} and \ref{sectwist}. Alternatively, the twisting rigidity can be allowed to vary in a manner consistent with the Ellipsoid model. We use a twist rigidity $B_3(\kappa_{\mathcal{L},3})$ which is curvature dependent to account for the variable twist rigidity of actin. We choose an arbitrary function~(\ref{formulaB3})
and vary the values of $\lambda_i$ to fit the data in Table~\ref{table:3}. The first line of the equation tends to $B_{\mathrm{min}}$ and $B_{\mathrm{max}}$ when $\kappa_{\mathcal{L},3} \rightarrow -\infty$ and $\kappa_{\mathcal{L},3} \rightarrow \infty$ respectively, while the second line of the equation tends to $0$ when $\kappa_{\mathcal{L},3} \rightarrow \pm \infty$. These properties ensure a good fit, with computationally simple inference, giving final parameters
\begin{eqnarray*}
&& \lambda_1 = 1.5928 \times 10^{-2}, \quad \lambda_2 = 7.2455 \times 10^{-3}, \quad \lambda_3 = 4.5982, \quad \lambda_4 = 1.0000 \times 10^{-6}, \\
&& \lambda_5 = 1.8196 \times 10^{-2}, \quad \lambda_6 = 4.5935 \times 10^{-3}, \quad \lambda_7 = 1.7203 \times 10^{-2}
\end{eqnarray*}
along with $B_{\mathrm{max}} = 1.6\times 10^4$~pN~$\mathrm{nm}^2$ and $B_{\mathrm{min}} = 9.6 \times 10^3$~pN~$\mathrm{nm}^2$ chosen as the rigidity limits that we reach for large curvature, with data and fitted curve seen in Figure~\ref{figure9}(a).

We also set the extension/compression rigidity by extending an ellipsoidal filament, fixing one end and applying a force on the other, and using the formula~(\ref{S3formula}). This has been done using the same set up as the torsion test above, with segments having their rotational orientation fixed. We also set a fixed extension in one of the end segments and measure the force after the system relaxes as an equivalent check. The estimated rigidity $S_3 = 3.4 \times 10^4$~pN is the third diagonal element of the bending rigidity matrix $\bm{S}$ for the Cosserat model. When compared to the relation $S_3 = E A \approx 1.3 \times 10^4$~pN, this suggests that Young's modulus $E$ or filament cross-section $A$ could be $2.7$ times larger than expected. Setting the final component of the shear matrix $S_2$ and $S_3$ is less straightforward from direct measurement, so we assume that the factor of $2.7$ stretch scaling translates directly into $S_1 = S_2 = 4 G A/3 \approx 1.6 \times 10^4$~pN. The rigidities used in the model for the Cosserat model are given in Table~\ref{table:4}, in addition to the scaled interface rigidities for the Ellipsoid model.   
\begin{table}[t]
\centering
\begin{tabular}{||c c | c ||} 
 \hline
 \rowcolor[HTML]{ccddff} \multicolumn{2}{||c|}{Parameter} & Value\\ [0.5ex] 
 \hline\hline
 Cosserat shearing rigidity  & $S_{1,2}$ & $1.6 \times 10^4$~pN\\
 Cosserat stretching rigidity & $S_{3}$ & $3.4 \times 10^4$~pN \\
 Cosserat bending rigidity & $B_{1,2}$ & $3.9 \times 10^4$~pN~$\mathrm{nm}^2$\\
 Cosserat twisting rigidity & $B_{3}$ & $1.0-1.6 \times 10^4$~pN~$\mathrm{nm}^2$ \\[0.5ex] 
 \hline 
 Ellipsoid lateral interface rigidity & $S_{\perp}$ & $2.3 \times 10^3$~pN\\
 Ellipsoid longitudinal interface rigidity & $S_{\parallel}$ & $3.4 \times 10^3$~pN\\[0.5ex]
 \hline
\end{tabular}
\caption{{\it Parameters used for the Ellipsoid and Cosserat model for multi-resolution model.}}
\label{table:4}
\end{table}

While the Cosserat model is able to capture the bending dynamics displayed by the Ellipsoid model, we have observed that it performs less well under torsional forces, with variable twisting rigidity. While the double helical nature of the filament does not cause torsion under simple bending and stretching, the filament extends and contracts under twists of varying magnitude. This suggests that rather than acting with forces, we may be able to instead directly relate the curvature to the rest length; if $\hat{\kappa}_{\mathcal{L},3}$ is large, the filament length would naturally be smaller or larger by some factor. In practice, this can be achieved in our discretisation by first calculating the curvature using the standard rest length of the Voronoi region given by equation (\ref{curvvector}). We then define function (\ref{polxikappa}) that varies in the twist curvature,
where $p_i$ are parameters to be inferred. We use  function (\ref{polxikappa}) to scale the extension quantities of the model related to the segments and Voronoi regions, labelled $e_i$ and $\mathcal{E}_i$ respectively. As we only know $\hat{\kappa}_{\mathcal{L},3}$ at discrete points (i.e. $[\hat{\kappa}_{\mathcal{L},3}]_j$ for $j=1,2,\dots,n-1$) over the course of the filament, we estimate the curvature of the segments as 
\begin{equation*}
    \hat{\kappa}^{\mathrm{seg}}_i = \mathcal{A}^h_i(\{[\hat{\kappa}_{\mathcal{L},3}]_j\}) \qquad \mbox{for} \; i=1,2,\dots,n \, ,
\end{equation*}
where $\mathcal{A}^h_i(\cdot)$ is the trapezoidal quadrature operator defined in equation~(\ref{trapoper}). We apply this to the extensions
\begin{equation*}
e_i = \frac{1}{\xi(\hat{\kappa}^{\mathrm{seg}}_i)} \cdot \frac{|\bm{l}_i|}{|\hat{\bm{l}}_i|}\, , \qquad \mathcal{E}_i = \frac{e_i + e_{i+1}}{2}, \qquad \mbox{for} \; i=1,2,\dots,n \, ,
\end{equation*}
and these quantities feed in directly to the normal acceleration calculations, meaning the Cosserat model will now stretch when twisted. We now need to establish parameters for equation~(\ref{polxikappa}) based on the behaviour of the Ellipsoid model under torsion. 
\begin{table}[t]
\centering
\begin{tabular}{||c | c | c ||c | c | c ||} 
 \hline
 \rowcolor[HTML]{ccddff}\multicolumn{3}{||c||}{Under-twist} &  \multicolumn{3}{|c||}{Over-twist}\\
 \hline\hline
 \rowcolor[HTML]{ccddff} $\phi$ & $\kappa_{\mathcal{L},3}$ & $\Delta L / L$ & $\phi$ & $\kappa_{\mathcal{L},3}$ & $\Delta L / L$\\ [0.5ex] 
 \hline
 $240^{\circ}$  & $-0.0555$ & $-0.0126$ & $240^{\circ}$  & $0.0555$ & $-0.0434$\\
 $220^{\circ}$  & $-0.0508$ & $-0.0108$ & $220^{\circ}$  & $0.0508$ & $-0.0378$\\
 \hline
 $200^{\circ}$  & $-0.0462$ & $-0.0090$ & $200^{\circ}$  & $0.0462$ & $-0.0341$\\
 
 $180^{\circ}$  & $-0.0416$ & $-0.0075$ & $180^{\circ}$  & $0.0416$ & $-0.0284$\\
 $160^{\circ}$  & $-0.0370$ & $-0.0057$ & $160^{\circ}$  & $0.0370$ & $-0.0256$\\
 $140^{\circ}$  & $-0.0324$ & $-0.0045$ & $140^{\circ}$  & $0.0324$ & $-0.0219$\\
 $120^{\circ}$  & $-0.0277$ & $-0.0032$ & $120^{\circ}$  & $0.0277$ & $-0.0186$\\
 $100^{\circ}$  & $-0.0231$ & $-0.0020$ & $100^{\circ}$  & $0.0231$ & $-0.0156$\\
 
 $80^{\circ}$  & $-0.0185$ & $-0.0009$ & $80^{\circ}$  & $0.0185$ & $-0.0121$\\
 $60^{\circ}$  & $-0.0139$ & $-0.0002$ & $60^{\circ}$  & $0.0139$ & $-0.0086$\\
 $40^{\circ}$  & $-0.0092$ & $0.0003$ & $40^{\circ}$  & $0.0092$ & $-0.0053$\\
 $20^{\circ}$  & $-0.0047$ & $0.0006$ & $20^{\circ}$  & $0.0047$ & $-0.0022$\\
 $10^{\circ}$  & $-0.0023$ & $0.0004$ & $10^{\circ}$  & $0.0023$ & $-0.0009$\\[0.5ex]
 \hline
\end{tabular}
\caption[Compression/extension of an ellipsoidal actin filament at various curvatures.]{{\it Compression/extension of an ellipsoidal actin filament at various curvatures.}}
\label{table:5}
\end{table}
In Table~\ref{table:5} we present data showing the change in filament length of the Ellipsoid model under torsion, and observe it is not symmetric. The more a filament is over-twisted, the shorter it gets. However, when it is under-twisted it initially lengthens, before slightly contracting, though at a smaller magnitude than when over-twisted. We use this data to infer the correct form of $\xi(\hat{\kappa}_{\mathcal{L},3}^i)$, with final parameters
\begin{eqnarray*}
&& p_6 = -2.6662 \times 10^{6}, \quad p_5 = -2.4927 \times 10^{4}, \quad p_4 = 1.0660 \times 10^{4}, \quad p_3 = 7.9376 \times 10^{1}, \\
&& p_2 = -2.0508 \times 10^{1}, \quad p_1 = -3.2416 \times 10^{-1}, \quad p_0 = -5.0666 \times 10^{-4}.
\end{eqnarray*}
with data and fitted curve for $\xi(\hat{\kappa}_{\mathcal{L},3}^i)$ presented in Figure~\ref{figure9}(b). It is this form of twist-stretch coupling that we employ for our final adapted twist example in Section~\ref{sectwist} presented in Figure~\ref{figure10}. 


\begin{thebibliography}{63}%
\makeatletter
\providecommand \@ifxundefined [1]{%
 \@ifx{#1\undefined}
}%
\providecommand \@ifnum [1]{%
 \ifnum #1\expandafter \@firstoftwo
 \else \expandafter \@secondoftwo
 \fi
}%
\providecommand \@ifx [1]{%
 \ifx #1\expandafter \@firstoftwo
 \else \expandafter \@secondoftwo
 \fi
}%
\providecommand \natexlab [1]{#1}%
\providecommand \enquote  [1]{``#1''}%
\providecommand \bibnamefont  [1]{#1}%
\providecommand \bibfnamefont [1]{#1}%
\providecommand \citenamefont [1]{#1}%
\providecommand \href@noop [0]{\@secondoftwo}%
\providecommand \href [0]{\begingroup \@sanitize@url \@href}%
\providecommand \@href[1]{\@@startlink{#1}\@@href}%
\providecommand \@@href[1]{\endgroup#1\@@endlink}%
\providecommand \@sanitize@url [0]{\catcode `\\12\catcode `\$12\catcode
  `\&12\catcode `\#12\catcode `\^12\catcode `\_12\catcode `\%12\relax}%
\providecommand \@@startlink[1]{}%
\providecommand \@@endlink[0]{}%
\providecommand \url  [0]{\begingroup\@sanitize@url \@url }%
\providecommand \@url [1]{\endgroup\@href {#1}{\urlprefix }}%
\providecommand \urlprefix  [0]{URL }%
\providecommand \Eprint [0]{\href }%
\providecommand \doibase [0]{http://dx.doi.org/}%
\providecommand \selectlanguage [0]{\@gobble}%
\providecommand \bibinfo  [0]{\@secondoftwo}%
\providecommand \bibfield  [0]{\@secondoftwo}%
\providecommand \translation [1]{[#1]}%
\providecommand \BibitemOpen [0]{}%
\providecommand \bibitemStop [0]{}%
\providecommand \bibitemNoStop [0]{.\EOS\space}%
\providecommand \EOS [0]{\spacefactor3000\relax}%
\providecommand \BibitemShut  [1]{\csname bibitem#1\endcsname}%
\let\auto@bib@innerbib\@empty
\bibitem [{\citenamefont {Floyd}\ \emph {et~al.}(2022)\citenamefont {Floyd},
  \citenamefont {Ni}, \citenamefont {Gunaratne}, \citenamefont {Erban},\ and\
  \citenamefont {Papoian}}]{Floyd:2022:SBSI}%
  \BibitemOpen
  \bibfield  {author} {\bibinfo {author} {\bibfnamefont {C.}~\bibnamefont
  {Floyd}}, \bibinfo {author} {\bibfnamefont {H.}~\bibnamefont {Ni}}, \bibinfo
  {author} {\bibfnamefont {R.}~\bibnamefont {Gunaratne}}, \bibinfo {author}
  {\bibfnamefont {R.}~\bibnamefont {Erban}}, \ and\ \bibinfo {author}
  {\bibfnamefont {G.~A.}\ \bibnamefont {Papoian}},\ }\href@noop {} {\enquote
  {\bibinfo {title} {On stretching, bending, shearing and twisting of actin
  filaments {I}: Variational models},}\ } (\bibinfo {year} {2022}),\ \bibinfo
  {note} {submitted to {\it Journal of Chemical Theory and Computation},
  available as https://arxiv.org/abs/2112.01480}\BibitemShut {NoStop}%
\bibitem [{\citenamefont {Lodish}\ \emph {et~al.}(2021)\citenamefont {Lodish},
  \citenamefont {Berk}, \citenamefont {C.}, \citenamefont {Krieger},
  \citenamefont {Bretscher}, \citenamefont {Ploegh}, \citenamefont {Martin},
  \citenamefont {Yaffe},\ and\ \citenamefont {Amon}}]{Lodish:2021:MCB}%
  \BibitemOpen
  \bibfield  {author} {\bibinfo {author} {\bibfnamefont {H.}~\bibnamefont
  {Lodish}}, \bibinfo {author} {\bibfnamefont {A.}~\bibnamefont {Berk}},
  \bibinfo {author} {\bibfnamefont {K.}~\bibnamefont {C.}}, \bibinfo {author}
  {\bibfnamefont {M.}~\bibnamefont {Krieger}}, \bibinfo {author} {\bibfnamefont
  {A.}~\bibnamefont {Bretscher}}, \bibinfo {author} {\bibfnamefont
  {H.}~\bibnamefont {Ploegh}}, \bibinfo {author} {\bibfnamefont
  {K.}~\bibnamefont {Martin}}, \bibinfo {author} {\bibfnamefont
  {M.}~\bibnamefont {Yaffe}}, \ and\ \bibinfo {author} {\bibfnamefont
  {A.}~\bibnamefont {Amon}},\ }\href@noop {} {\emph {\bibinfo {title}
  {Molecular Cell Biology}}},\ \bibinfo {edition} {9th}\ ed.\ (\bibinfo {year}
  {2021})\BibitemShut {NoStop}%
\bibitem [{\citenamefont {Dominguez}\ and\ \citenamefont
  {Holmes}(2011)}]{Dominguez2011ActinFunction}%
  \BibitemOpen
  \bibfield  {author} {\bibinfo {author} {\bibfnamefont {R.}~\bibnamefont
  {Dominguez}}\ and\ \bibinfo {author} {\bibfnamefont {K.~C.}\ \bibnamefont
  {Holmes}},\ }\href {\doibase 10.7312/blis92798-008} {\bibfield  {journal}
  {\bibinfo  {journal} {Annu Rev Biophys}\ ,\ \bibinfo {pages} {169}} (\bibinfo
  {year} {2011})}\BibitemShut {NoStop}%
\bibitem [{\citenamefont {Merino}\ \emph {et~al.}(2018)\citenamefont {Merino},
  \citenamefont {Pospich}, \citenamefont {Funk}, \citenamefont {Wagner},
  \citenamefont {K{\"{u}}llmer}, \citenamefont {Arndt}, \citenamefont
  {Bieling},\ and\ \citenamefont {Raunser}}]{Merino2018StructuralCryo-EM}%
  \BibitemOpen
  \bibfield  {author} {\bibinfo {author} {\bibfnamefont {F.}~\bibnamefont
  {Merino}}, \bibinfo {author} {\bibfnamefont {S.}~\bibnamefont {Pospich}},
  \bibinfo {author} {\bibfnamefont {J.}~\bibnamefont {Funk}}, \bibinfo {author}
  {\bibfnamefont {T.}~\bibnamefont {Wagner}}, \bibinfo {author} {\bibfnamefont
  {F.}~\bibnamefont {K{\"{u}}llmer}}, \bibinfo {author} {\bibfnamefont {H.~D.}\
  \bibnamefont {Arndt}}, \bibinfo {author} {\bibfnamefont {P.}~\bibnamefont
  {Bieling}}, \ and\ \bibinfo {author} {\bibfnamefont {S.}~\bibnamefont
  {Raunser}},\ }\href {\doibase 10.1038/s41594-018-0074-0} {\bibfield
  {journal} {\bibinfo  {journal} {Nature Structural and Molecular Biology}\
  }\textbf {\bibinfo {volume} {25}},\ \bibinfo {pages} {528} (\bibinfo {year}
  {2018})}\BibitemShut {NoStop}%
\bibitem [{\citenamefont {Chou}\ and\ \citenamefont
  {Pollard}(2019)}]{Chou2019MechanismNucleotides}%
  \BibitemOpen
  \bibfield  {author} {\bibinfo {author} {\bibfnamefont {S.~Z.}\ \bibnamefont
  {Chou}}\ and\ \bibinfo {author} {\bibfnamefont {T.~D.}\ \bibnamefont
  {Pollard}},\ }\href {\doibase 10.1073/pnas.1807028115} {\bibfield  {journal}
  {\bibinfo  {journal} {Proceedings of the National Academy of Sciences of the
  United States of America}\ }\textbf {\bibinfo {volume} {116}},\ \bibinfo
  {pages} {4265} (\bibinfo {year} {2019})}\BibitemShut {NoStop}%
\bibitem [{\citenamefont {Berman}\ \emph {et~al.}(2000)\citenamefont {Berman},
  \citenamefont {Westbrook}, \citenamefont {Feng}, \citenamefont {Gilliland},
  \citenamefont {Bhat}, \citenamefont {Weissig}, \citenamefont {Shindyalov},\
  and\ \citenamefont {Bourne}}]{Berman2000TheBank}%
  \BibitemOpen
  \bibfield  {author} {\bibinfo {author} {\bibfnamefont {H.~M.}\ \bibnamefont
  {Berman}}, \bibinfo {author} {\bibfnamefont {J.}~\bibnamefont {Westbrook}},
  \bibinfo {author} {\bibfnamefont {Z.}~\bibnamefont {Feng}}, \bibinfo {author}
  {\bibfnamefont {G.}~\bibnamefont {Gilliland}}, \bibinfo {author}
  {\bibfnamefont {T.~N.}\ \bibnamefont {Bhat}}, \bibinfo {author}
  {\bibfnamefont {H.}~\bibnamefont {Weissig}}, \bibinfo {author} {\bibfnamefont
  {I.~N.}\ \bibnamefont {Shindyalov}}, \ and\ \bibinfo {author} {\bibfnamefont
  {P.~E.}\ \bibnamefont {Bourne}},\ }\href
  {http://www.rcsb.org/pdb/status.html} {\emph {\bibinfo {title} {Nucleic Acids
  Research}}},\ \bibinfo {type} {Tech. Rep.}\ \bibinfo {number} {1}\ (\bibinfo
  {year} {2000})\BibitemShut {NoStop}%
\bibitem [{\citenamefont {Pfaendtner}\ \emph {et~al.}(2010)\citenamefont
  {Pfaendtner}, \citenamefont {Lyman}, \citenamefont {Pollard},\ and\
  \citenamefont {Voth}}]{Pfaendtner2010StructureFilament}%
  \BibitemOpen
  \bibfield  {author} {\bibinfo {author} {\bibfnamefont {J.}~\bibnamefont
  {Pfaendtner}}, \bibinfo {author} {\bibfnamefont {E.}~\bibnamefont {Lyman}},
  \bibinfo {author} {\bibfnamefont {T.~D.}\ \bibnamefont {Pollard}}, \ and\
  \bibinfo {author} {\bibfnamefont {G.~A.}\ \bibnamefont {Voth}},\ }\href
  {\doibase 10.1016/j.jmb.2009.11.034} {\bibfield  {journal} {\bibinfo
  {journal} {Journal of Molecular Biology}\ }\textbf {\bibinfo {volume}
  {396}},\ \bibinfo {pages} {252} (\bibinfo {year} {2010})}\BibitemShut
  {NoStop}%
\bibitem [{\citenamefont {Shamloo}\ and\ \citenamefont
  {Mehrafrooz}(2018)}]{Shamloo2018NanomechanicsSimulation}%
  \BibitemOpen
  \bibfield  {author} {\bibinfo {author} {\bibfnamefont {A.}~\bibnamefont
  {Shamloo}}\ and\ \bibinfo {author} {\bibfnamefont {B.}~\bibnamefont
  {Mehrafrooz}},\ }\href {\doibase 10.1002/cm.21429} {\bibfield  {journal}
  {\bibinfo  {journal} {Cytoskeleton}\ }\textbf {\bibinfo {volume} {75}},\
  \bibinfo {pages} {118} (\bibinfo {year} {2018})}\BibitemShut {NoStop}%
\bibitem [{\citenamefont {Tanaka}\ \emph {et~al.}(2018)\citenamefont {Tanaka},
  \citenamefont {Takeda}, \citenamefont {Mitsuoka}, \citenamefont {Oda},
  \citenamefont {Kimura-Sakiyama}, \citenamefont {Ma{\'{e}}da},\ and\
  \citenamefont {Narita}}]{Tanaka2018StructuralDisassembly}%
  \BibitemOpen
  \bibfield  {author} {\bibinfo {author} {\bibfnamefont {K.}~\bibnamefont
  {Tanaka}}, \bibinfo {author} {\bibfnamefont {S.}~\bibnamefont {Takeda}},
  \bibinfo {author} {\bibfnamefont {K.}~\bibnamefont {Mitsuoka}}, \bibinfo
  {author} {\bibfnamefont {T.}~\bibnamefont {Oda}}, \bibinfo {author}
  {\bibfnamefont {C.}~\bibnamefont {Kimura-Sakiyama}}, \bibinfo {author}
  {\bibfnamefont {Y.}~\bibnamefont {Ma{\'{e}}da}}, \ and\ \bibinfo {author}
  {\bibfnamefont {A.}~\bibnamefont {Narita}},\ }\href {\doibase
  10.1038/s41467-018-04290-w} {\bibfield  {journal} {\bibinfo  {journal}
  {Nature Communications}\ }\textbf {\bibinfo {volume} {9}} (\bibinfo {year}
  {2018}),\ 10.1038/s41467-018-04290-w}\BibitemShut {NoStop}%
\bibitem [{\citenamefont {Narita}\ \emph {et~al.}(2006)\citenamefont {Narita},
  \citenamefont {Takeda}, \citenamefont {Yamashita},\ and\ \citenamefont
  {Ma{\'{e}}da}}]{Narita2006StructuralStudy}%
  \BibitemOpen
  \bibfield  {author} {\bibinfo {author} {\bibfnamefont {A.}~\bibnamefont
  {Narita}}, \bibinfo {author} {\bibfnamefont {S.}~\bibnamefont {Takeda}},
  \bibinfo {author} {\bibfnamefont {A.}~\bibnamefont {Yamashita}}, \ and\
  \bibinfo {author} {\bibfnamefont {Y.}~\bibnamefont {Ma{\'{e}}da}},\ }\href
  {\doibase 10.1038/sj.emboj.7601395} {\bibfield  {journal} {\bibinfo
  {journal} {EMBO Journal}\ }\textbf {\bibinfo {volume} {25}},\ \bibinfo
  {pages} {5626} (\bibinfo {year} {2006})}\BibitemShut {NoStop}%
\bibitem [{\citenamefont {Oda}\ \emph {et~al.}(2009)\citenamefont {Oda},
  \citenamefont {Iwasa}, \citenamefont {Aihara}, \citenamefont {Ma{\'{e}}da},\
  and\ \citenamefont {Narita}}]{Oda2009TheTransition}%
  \BibitemOpen
  \bibfield  {author} {\bibinfo {author} {\bibfnamefont {T.}~\bibnamefont
  {Oda}}, \bibinfo {author} {\bibfnamefont {M.}~\bibnamefont {Iwasa}}, \bibinfo
  {author} {\bibfnamefont {T.}~\bibnamefont {Aihara}}, \bibinfo {author}
  {\bibfnamefont {Y.}~\bibnamefont {Ma{\'{e}}da}}, \ and\ \bibinfo {author}
  {\bibfnamefont {A.}~\bibnamefont {Narita}},\ }\href {\doibase
  10.1038/nature07685} {\bibfield  {journal} {\bibinfo  {journal} {Nature}\
  }\textbf {\bibinfo {volume} {457}},\ \bibinfo {pages} {441} (\bibinfo {year}
  {2009})}\BibitemShut {NoStop}%
\bibitem [{\citenamefont {Chu}\ and\ \citenamefont
  {Voth}(2005)}]{Chu2005AllosteryAnalysis}%
  \BibitemOpen
  \bibfield  {author} {\bibinfo {author} {\bibfnamefont {J.~W.}\ \bibnamefont
  {Chu}}\ and\ \bibinfo {author} {\bibfnamefont {G.~A.}\ \bibnamefont {Voth}},\
  }\href {\doibase 10.1073/pnas.0503732102} {\bibfield  {journal} {\bibinfo
  {journal} {Proceedings of the National Academy of Sciences of the United
  States of America}\ }\textbf {\bibinfo {volume} {102}},\ \bibinfo {pages}
  {13111} (\bibinfo {year} {2005})}\BibitemShut {NoStop}%
\bibitem [{\citenamefont {Dobramysl}\ \emph {et~al.}(2016)\citenamefont
  {Dobramysl}, \citenamefont {Papoian},\ and\ \citenamefont
  {Erban}}]{Dobramysl2016StericFilopodia}%
  \BibitemOpen
  \bibfield  {author} {\bibinfo {author} {\bibfnamefont {U.}~\bibnamefont
  {Dobramysl}}, \bibinfo {author} {\bibfnamefont {G.~A.}\ \bibnamefont
  {Papoian}}, \ and\ \bibinfo {author} {\bibfnamefont {R.}~\bibnamefont
  {Erban}},\ }\href {\doibase 10.1016/j.bpj.2016.03.013} {\bibfield  {journal}
  {\bibinfo  {journal} {Biophysical Journal}\ }\textbf {\bibinfo {volume}
  {110}},\ \bibinfo {pages} {2066} (\bibinfo {year} {2016})}\BibitemShut
  {NoStop}%
\bibitem [{\citenamefont {Schramm}\ \emph
  {et~al.}(2017{\natexlab{a}})\citenamefont {Schramm}, \citenamefont {Hocky},
  \citenamefont {Voth}, \citenamefont {Blanchoin}, \citenamefont {Martiel},\
  and\ \citenamefont {De~La~Cruz}}]{Schramm2017ActinDissociation}%
  \BibitemOpen
  \bibfield  {author} {\bibinfo {author} {\bibfnamefont {A.~C.}\ \bibnamefont
  {Schramm}}, \bibinfo {author} {\bibfnamefont {G.~M.}\ \bibnamefont {Hocky}},
  \bibinfo {author} {\bibfnamefont {G.~A.}\ \bibnamefont {Voth}}, \bibinfo
  {author} {\bibfnamefont {L.}~\bibnamefont {Blanchoin}}, \bibinfo {author}
  {\bibfnamefont {J.~L.}\ \bibnamefont {Martiel}}, \ and\ \bibinfo {author}
  {\bibfnamefont {E.~M.}\ \bibnamefont {De~La~Cruz}},\ }\href {\doibase
  10.1016/j.bpj.2017.05.016} {\bibfield  {journal} {\bibinfo  {journal}
  {Biophysical Journal}\ }\textbf {\bibinfo {volume} {112}},\ \bibinfo {pages}
  {2624} (\bibinfo {year} {2017}{\natexlab{a}})}\BibitemShut {NoStop}%
\bibitem [{\citenamefont {Schramm}\ \emph
  {et~al.}(2019{\natexlab{a}})\citenamefont {Schramm}, \citenamefont {Hocky},
  \citenamefont {Voth}, \citenamefont {Martiel},\ and\ \citenamefont
  {De~La~Cruz}}]{Schramm2019PlasticFilaments}%
  \BibitemOpen
  \bibfield  {author} {\bibinfo {author} {\bibfnamefont {A.~C.}\ \bibnamefont
  {Schramm}}, \bibinfo {author} {\bibfnamefont {G.~M.}\ \bibnamefont {Hocky}},
  \bibinfo {author} {\bibfnamefont {G.~A.}\ \bibnamefont {Voth}}, \bibinfo
  {author} {\bibfnamefont {J.~L.}\ \bibnamefont {Martiel}}, \ and\ \bibinfo
  {author} {\bibfnamefont {E.~M.}\ \bibnamefont {De~La~Cruz}},\ }\href
  {\doibase 10.1016/j.bpj.2019.06.018} {\bibfield  {journal} {\bibinfo
  {journal} {Biophysical Journal}\ }\textbf {\bibinfo {volume} {117}},\
  \bibinfo {pages} {453} (\bibinfo {year} {2019}{\natexlab{a}})}\BibitemShut
  {NoStop}%
\bibitem [{\citenamefont {Chu}\ and\ \citenamefont
  {Voth}(2006)}]{Chu2006Coarse-grainedSimulations}%
  \BibitemOpen
  \bibfield  {author} {\bibinfo {author} {\bibfnamefont {J.~W.}\ \bibnamefont
  {Chu}}\ and\ \bibinfo {author} {\bibfnamefont {G.~A.}\ \bibnamefont {Voth}},\
  }\href {\doibase 10.1529/biophysj.105.073924} {\bibfield  {journal} {\bibinfo
   {journal} {Biophysical Journal}\ }\textbf {\bibinfo {volume} {90}},\
  \bibinfo {pages} {1572} (\bibinfo {year} {2006})}\BibitemShut {NoStop}%
\bibitem [{\citenamefont {Horan}\ and\ \citenamefont
  {Vavylonis}(2019)}]{Horan2019InsightsModel}%
  \BibitemOpen
  \bibfield  {author} {\bibinfo {author} {\bibfnamefont {B.}~\bibnamefont
  {Horan}}\ and\ \bibinfo {author} {\bibfnamefont {D.}~\bibnamefont
  {Vavylonis}},\ }\href {\doibase 10.1101/715383} {\bibfield  {journal}
  {\bibinfo  {journal} {bioRxiv}\ ,\ \bibinfo {pages} {715383}} (\bibinfo
  {year} {2019})}\BibitemShut {NoStop}%
\bibitem [{\citenamefont {Katkar}\ \emph {et~al.}(2018)\citenamefont {Katkar},
  \citenamefont {Davtyan}, \citenamefont {Durumeric}, \citenamefont {Hocky},
  \citenamefont {Schramm}, \citenamefont {De~La~Cruz},\ and\ \citenamefont
  {Voth}}]{Katkar2018InsightsFilaments}%
  \BibitemOpen
  \bibfield  {author} {\bibinfo {author} {\bibfnamefont {H.~H.}\ \bibnamefont
  {Katkar}}, \bibinfo {author} {\bibfnamefont {A.}~\bibnamefont {Davtyan}},
  \bibinfo {author} {\bibfnamefont {A.~E.}\ \bibnamefont {Durumeric}}, \bibinfo
  {author} {\bibfnamefont {G.~M.}\ \bibnamefont {Hocky}}, \bibinfo {author}
  {\bibfnamefont {A.~C.}\ \bibnamefont {Schramm}}, \bibinfo {author}
  {\bibfnamefont {E.~M.}\ \bibnamefont {De~La~Cruz}}, \ and\ \bibinfo {author}
  {\bibfnamefont {G.~A.}\ \bibnamefont {Voth}},\ }\href {\doibase
  10.1016/j.bpj.2018.08.034} {\bibfield  {journal} {\bibinfo  {journal}
  {Biophysical Journal}\ }\textbf {\bibinfo {volume} {115}},\ \bibinfo {pages}
  {1589} (\bibinfo {year} {2018})}\BibitemShut {NoStop}%
\bibitem [{\citenamefont {Savelyev}\ and\ \citenamefont
  {Papoian}(2010)}]{Savelyev2010ChemicallyDNA}%
  \BibitemOpen
  \bibfield  {author} {\bibinfo {author} {\bibfnamefont {A.}~\bibnamefont
  {Savelyev}}\ and\ \bibinfo {author} {\bibfnamefont {G.~A.}\ \bibnamefont
  {Papoian}},\ }\href {\doibase 10.1073/pnas.1001163107} {\bibfield  {journal}
  {\bibinfo  {journal} {Proceedings of the National Academy of Sciences of the
  United States of America}\ }\textbf {\bibinfo {volume} {107}},\ \bibinfo
  {pages} {20340} (\bibinfo {year} {2010})}\BibitemShut {NoStop}%
\bibitem [{\citenamefont {Aydin}\ \emph {et~al.}(2018)\citenamefont {Aydin},
  \citenamefont {Katkar},\ and\ \citenamefont
  {Voth}}]{Aydin2018MultiscaleProteins}%
  \BibitemOpen
  \bibfield  {author} {\bibinfo {author} {\bibfnamefont {F.}~\bibnamefont
  {Aydin}}, \bibinfo {author} {\bibfnamefont {H.~H.}\ \bibnamefont {Katkar}}, \
  and\ \bibinfo {author} {\bibfnamefont {G.~A.}\ \bibnamefont {Voth}},\ }\href
  {\doibase 10.1007/s12551-018-0474-8} {\enquote {\bibinfo {title} {{Multiscale
  simulation of actin filaments and actin-associated proteins}},}\ } (\bibinfo
  {year} {2018})\BibitemShut {NoStop}%
\bibitem [{\citenamefont {Yamaoka}\ \emph {et~al.}(2012)\citenamefont
  {Yamaoka}, \citenamefont {Matsushita}, \citenamefont {Shimada},\ and\
  \citenamefont {Adachi}}]{Yamaoka2012MultiscaleCytoskeleton}%
  \BibitemOpen
  \bibfield  {author} {\bibinfo {author} {\bibfnamefont {H.}~\bibnamefont
  {Yamaoka}}, \bibinfo {author} {\bibfnamefont {S.}~\bibnamefont {Matsushita}},
  \bibinfo {author} {\bibfnamefont {Y.}~\bibnamefont {Shimada}}, \ and\
  \bibinfo {author} {\bibfnamefont {T.}~\bibnamefont {Adachi}},\ }\href
  {\doibase 10.1007/s10237-011-0317-z} {\enquote {\bibinfo {title} {{Multiscale
  modeling and mechanics of filamentous actin cytoskeleton}},}\ } (\bibinfo
  {year} {2012})\BibitemShut {NoStop}%
\bibitem [{\citenamefont {De~La~Cruz}\ \emph {et~al.}(2010)\citenamefont
  {De~La~Cruz}, \citenamefont {Roland}, \citenamefont {McCullough},
  \citenamefont {Blanchoin},\ and\ \citenamefont
  {Martiel}}]{DeLaCruz2010OriginFilaments}%
  \BibitemOpen
  \bibfield  {author} {\bibinfo {author} {\bibfnamefont {E.~M.}\ \bibnamefont
  {De~La~Cruz}}, \bibinfo {author} {\bibfnamefont {J.}~\bibnamefont {Roland}},
  \bibinfo {author} {\bibfnamefont {B.~R.}\ \bibnamefont {McCullough}},
  \bibinfo {author} {\bibfnamefont {L.}~\bibnamefont {Blanchoin}}, \ and\
  \bibinfo {author} {\bibfnamefont {J.~L.}\ \bibnamefont {Martiel}},\ }\href
  {\doibase 10.1016/j.bpj.2010.07.009} {\bibfield  {journal} {\bibinfo
  {journal} {Biophysical Journal}\ }\textbf {\bibinfo {volume} {99}},\ \bibinfo
  {pages} {1852} (\bibinfo {year} {2010})}\BibitemShut {NoStop}%
\bibitem [{\citenamefont {Fogarty}\ \emph {et~al.}(2016)\citenamefont
  {Fogarty}, \citenamefont {Potestio},\ and\ \citenamefont
  {Kremer}}]{Fogarty2016ASite}%
  \BibitemOpen
  \bibfield  {author} {\bibinfo {author} {\bibfnamefont {A.~C.}\ \bibnamefont
  {Fogarty}}, \bibinfo {author} {\bibfnamefont {R.}~\bibnamefont {Potestio}}, \
  and\ \bibinfo {author} {\bibfnamefont {K.}~\bibnamefont {Kremer}},\ }\href
  {\doibase 10.1002/prot.25173} {\bibfield  {journal} {\bibinfo  {journal}
  {Proteins: Structure, Function and Bioinformatics}\ }\textbf {\bibinfo
  {volume} {84}},\ \bibinfo {pages} {1902} (\bibinfo {year}
  {2016})}\BibitemShut {NoStop}%
\bibitem [{\citenamefont {Rolls}\ \emph {et~al.}(2016)\citenamefont {Rolls},
  \citenamefont {Togashi},\ and\ \citenamefont
  {Erban}}]{Rolls2016VaryingDynamics}%
  \BibitemOpen
  \bibfield  {author} {\bibinfo {author} {\bibfnamefont {E.}~\bibnamefont
  {Rolls}}, \bibinfo {author} {\bibfnamefont {Y.}~\bibnamefont {Togashi}}, \
  and\ \bibinfo {author} {\bibfnamefont {R.}~\bibnamefont {Erban}},\ }\href
  {http://arxiv.org/abs/1607.08062} {\bibfield  {journal} {\bibinfo  {journal}
  {To appear in Multiscale Modelling and Simulation}\ } (\bibinfo {year}
  {2016})}\BibitemShut {NoStop}%
\bibitem [{\citenamefont {Rolls}\ and\ \citenamefont
  {Erban}(2018)}]{Rolls2018Multi-resolutionInteractions}%
  \BibitemOpen
  \bibfield  {author} {\bibinfo {author} {\bibfnamefont {E.}~\bibnamefont
  {Rolls}}\ and\ \bibinfo {author} {\bibfnamefont {R.}~\bibnamefont {Erban}},\
  }\href {\doibase 10.1063/1.5018595} {\bibfield  {journal} {\bibinfo
  {journal} {Journal of Chemical Physics}\ }\textbf {\bibinfo {volume} {148}}
  (\bibinfo {year} {2018}),\ 10.1063/1.5018595}\BibitemShut {NoStop}%
\bibitem [{\citenamefont {Zavadlav}\ \emph {et~al.}(2015)\citenamefont
  {Zavadlav}, \citenamefont {Podgornik},\ and\ \citenamefont
  {Praprotnik}}]{Zavadlav2015AdaptiveSolution}%
  \BibitemOpen
  \bibfield  {author} {\bibinfo {author} {\bibfnamefont {J.}~\bibnamefont
  {Zavadlav}}, \bibinfo {author} {\bibfnamefont {R.}~\bibnamefont {Podgornik}},
  \ and\ \bibinfo {author} {\bibfnamefont {M.}~\bibnamefont {Praprotnik}},\
  }\href {\doibase 10.1021/acs.jctc.5b00596} {\bibfield  {journal} {\bibinfo
  {journal} {Journal of Chemical Theory and Computation}\ }\textbf {\bibinfo
  {volume} {11}},\ \bibinfo {pages} {5035} (\bibinfo {year}
  {2015})}\BibitemShut {NoStop}%
\bibitem [{\citenamefont {Erban}\ and\ \citenamefont
  {Chapman}(2020)}]{Chapman}%
  \BibitemOpen
  \bibfield  {author} {\bibinfo {author} {\bibfnamefont {R.}~\bibnamefont
  {Erban}}\ and\ \bibinfo {author} {\bibfnamefont {S.}~\bibnamefont
  {Chapman}},\ }\href@noop {} {\emph {\bibinfo {title} {Stochastic Modelling of
  Reaction–Diffusion Processes}}}\ (\bibinfo  {publisher} {Cambridge
  University Press},\ \bibinfo {year} {2020})\BibitemShut {NoStop}%
\bibitem [{\citenamefont {Praprotnik}\ \emph {et~al.}(2007)\citenamefont
  {Praprotnik}, \citenamefont {Delle~Site},\ and\ \citenamefont
  {Kremer}}]{Praprotnik2007ASimulation}%
  \BibitemOpen
  \bibfield  {author} {\bibinfo {author} {\bibfnamefont {M.}~\bibnamefont
  {Praprotnik}}, \bibinfo {author} {\bibfnamefont {L.}~\bibnamefont
  {Delle~Site}}, \ and\ \bibinfo {author} {\bibfnamefont {K.}~\bibnamefont
  {Kremer}},\ }\href {\doibase 10.1063/1.2714540} {\bibfield  {journal}
  {\bibinfo  {journal} {Journal of Chemical Physics}\ }\textbf {\bibinfo
  {volume} {126}} (\bibinfo {year} {2007}),\ 10.1063/1.2714540}\BibitemShut
  {NoStop}%
\bibitem [{\citenamefont {Erban}(2014)}]{Erban2014FromDynamics}%
  \BibitemOpen
  \bibfield  {author} {\bibinfo {author} {\bibfnamefont {R.}~\bibnamefont
  {Erban}},\ }\href {\doibase 10.1098/rspa.2014.0036} {\bibfield  {journal}
  {\bibinfo  {journal} {Proceedings of the Royal Society A: Mathematical,
  Physical and Engineering Sciences}\ }\textbf {\bibinfo {volume} {470}}
  (\bibinfo {year} {2014}),\ 10.1098/rspa.2014.0036}\BibitemShut {NoStop}%
\bibitem [{\citenamefont {Gunaratne}\ \emph {et~al.}(2019)\citenamefont
  {Gunaratne}, \citenamefont {Wilson}, \citenamefont {Flegg},\ and\
  \citenamefont {Erban}}]{Gunaratne2019Multi-resolutionInteractions}%
  \BibitemOpen
  \bibfield  {author} {\bibinfo {author} {\bibfnamefont {R.}~\bibnamefont
  {Gunaratne}}, \bibinfo {author} {\bibfnamefont {D.}~\bibnamefont {Wilson}},
  \bibinfo {author} {\bibfnamefont {M.}~\bibnamefont {Flegg}}, \ and\ \bibinfo
  {author} {\bibfnamefont {R.}~\bibnamefont {Erban}},\ }\href
  {http://arxiv.org/abs/1811.03484} {\bibfield  {journal} {\bibinfo  {journal}
  {Interface Focus (accepted)}\ } (\bibinfo {year} {2019})}\BibitemShut
  {NoStop}%
\bibitem [{\citenamefont {Erban}(2016)}]{Erban2016CouplingDynamics}%
  \BibitemOpen
  \bibfield  {author} {\bibinfo {author} {\bibfnamefont {R.}~\bibnamefont
  {Erban}},\ }\href {\doibase 10.1098/rspa.2015.0556} {\bibfield  {journal}
  {\bibinfo  {journal} {Proceedings of the Royal Society A: Mathematical,
  Physical and Engineering Sciences}\ }\textbf {\bibinfo {volume} {472}}
  (\bibinfo {year} {2016}),\ 10.1098/rspa.2015.0556}\BibitemShut {NoStop}%
\bibitem [{\citenamefont
  {Erban}(2019)}]{Erban2019Coarse-grainingDistributions}%
  \BibitemOpen
  \bibfield  {author} {\bibinfo {author} {\bibfnamefont {R.}~\bibnamefont
  {Erban}},\ }\href {\doibase 10.1007/s00285-019-01433-5} {\bibfield  {journal}
  {\bibinfo  {journal} {Journal of Mathematical Biology}\ } (\bibinfo {year}
  {2019}),\ 10.1007/s00285-019-01433-5}\BibitemShut {NoStop}%
\bibitem [{\citenamefont {Wang}\ \emph {et~al.}(2009)\citenamefont {Wang},
  \citenamefont {Noid}, \citenamefont {Liu},\ and\ \citenamefont
  {Voth}}]{Wang:2009:EFC}%
  \BibitemOpen
  \bibfield  {author} {\bibinfo {author} {\bibfnamefont {Y.}~\bibnamefont
  {Wang}}, \bibinfo {author} {\bibfnamefont {W.}~\bibnamefont {Noid}}, \bibinfo
  {author} {\bibfnamefont {P.}~\bibnamefont {Liu}}, \ and\ \bibinfo {author}
  {\bibfnamefont {G.}~\bibnamefont {Voth}},\ }\href@noop {} {\bibfield
  {journal} {\bibinfo  {journal} {Physical Chemistry Chemical Physics}\
  }\textbf {\bibinfo {volume} {11}},\ \bibinfo {pages} {2002} (\bibinfo {year}
  {2009})}\BibitemShut {NoStop}%
\bibitem [{\citenamefont {Joshi}\ and\ \citenamefont
  {Deshmukh}(2021)}]{Joshi:2020:RAC}%
  \BibitemOpen
  \bibfield  {author} {\bibinfo {author} {\bibfnamefont {S.}~\bibnamefont
  {Joshi}}\ and\ \bibinfo {author} {\bibfnamefont {S.}~\bibnamefont
  {Deshmukh}},\ }\href@noop {} {\bibfield  {journal} {\bibinfo  {journal}
  {Molecular Simulation}\ }\textbf {\bibinfo {volume} {47}},\ \bibinfo {pages}
  {786} (\bibinfo {year} {2021})}\BibitemShut {NoStop}%
\bibitem [{\citenamefont {Utterson}\ and\ \citenamefont
  {Erban}(2022)}]{Utterson:2022:SMF}%
  \BibitemOpen
  \bibfield  {author} {\bibinfo {author} {\bibfnamefont {J.}~\bibnamefont
  {Utterson}}\ and\ \bibinfo {author} {\bibfnamefont {R.}~\bibnamefont
  {Erban}},\ }\href@noop {} {\bibfield  {journal} {\bibinfo  {journal}
  {Physical Chemistry Chemical Physics}\ } (\bibinfo {year}
  {2022})}\BibitemShut {NoStop}%
\bibitem [{\citenamefont {Popov}\ \emph {et~al.}(2016)\citenamefont {Popov},
  \citenamefont {Komianos},\ and\ \citenamefont
  {Papoian}}]{Popov2016MEDYAN:Networks}%
  \BibitemOpen
  \bibfield  {author} {\bibinfo {author} {\bibfnamefont {K.}~\bibnamefont
  {Popov}}, \bibinfo {author} {\bibfnamefont {J.}~\bibnamefont {Komianos}}, \
  and\ \bibinfo {author} {\bibfnamefont {G.~A.}\ \bibnamefont {Papoian}},\
  }\href {\doibase 10.1371/journal.pcbi.1004877} {\bibfield  {journal}
  {\bibinfo  {journal} {PLoS Computational Biology}\ }\textbf {\bibinfo
  {volume} {12}},\ \bibinfo {pages} {e1004877} (\bibinfo {year}
  {2016})}\BibitemShut {NoStop}%
\bibitem [{\citenamefont {Floyd}\ \emph {et~al.}(2019)\citenamefont {Floyd},
  \citenamefont {Papoian},\ and\ \citenamefont
  {Jarzynski}}]{Floyd2019QuantifyingNetworks}%
  \BibitemOpen
  \bibfield  {author} {\bibinfo {author} {\bibfnamefont {C.}~\bibnamefont
  {Floyd}}, \bibinfo {author} {\bibfnamefont {G.~A.}\ \bibnamefont {Papoian}},
  \ and\ \bibinfo {author} {\bibfnamefont {C.}~\bibnamefont {Jarzynski}},\
  }\href {\doibase 10.1098/rsfs.2018.0078} {\bibfield  {journal} {\bibinfo
  {journal} {Interface Focus}\ }\textbf {\bibinfo {volume} {9}},\ \bibinfo
  {pages} {20180078} (\bibinfo {year} {2019})}\BibitemShut {NoStop}%
\bibitem [{\citenamefont {MacKintosh}\ \emph {et~al.}(1995)\citenamefont
  {MacKintosh}, \citenamefont {K{\"{a}}s},\ and\ \citenamefont
  {Janmey}}]{MacKintosh1995ElasticityNetworks}%
  \BibitemOpen
  \bibfield  {author} {\bibinfo {author} {\bibfnamefont {F.~C.}\ \bibnamefont
  {MacKintosh}}, \bibinfo {author} {\bibfnamefont {J.}~\bibnamefont
  {K{\"{a}}s}}, \ and\ \bibinfo {author} {\bibfnamefont {P.~A.}\ \bibnamefont
  {Janmey}},\ }\href {\doibase 10.1103/PhysRevLett.75.4425} {\bibfield
  {journal} {\bibinfo  {journal} {Physical Review Letters}\ }\textbf {\bibinfo
  {volume} {75}},\ \bibinfo {pages} {4425} (\bibinfo {year}
  {1995})}\BibitemShut {NoStop}%
\bibitem [{\citenamefont {Satcher}\ and\ \citenamefont
  {Dewey}(1996)}]{Satcher1996TheoreticalCytoskeleton}%
  \BibitemOpen
  \bibfield  {author} {\bibinfo {author} {\bibfnamefont {R.~L.}\ \bibnamefont
  {Satcher}}\ and\ \bibinfo {author} {\bibfnamefont {C.~F.}\ \bibnamefont
  {Dewey}},\ }\href {\doibase 10.1016/S0006-3495(96)79206-8} {\bibfield
  {journal} {\bibinfo  {journal} {Biophysical Journal}\ }\textbf {\bibinfo
  {volume} {71}},\ \bibinfo {pages} {109} (\bibinfo {year} {1996})}\BibitemShut
  {NoStop}%
\bibitem [{\citenamefont {Gazzola}\ \emph {et~al.}(2018)\citenamefont
  {Gazzola}, \citenamefont {Dudte}, \citenamefont {McCormick},\ and\
  \citenamefont {Mahadevan}}]{Gazzola2018ForwardFilaments}%
  \BibitemOpen
  \bibfield  {author} {\bibinfo {author} {\bibfnamefont {M.}~\bibnamefont
  {Gazzola}}, \bibinfo {author} {\bibfnamefont {L.~H.}\ \bibnamefont {Dudte}},
  \bibinfo {author} {\bibfnamefont {A.~G.}\ \bibnamefont {McCormick}}, \ and\
  \bibinfo {author} {\bibfnamefont {L.}~\bibnamefont {Mahadevan}},\ }\href
  {\doibase 10.1098/rsos.171628} {\bibfield  {journal} {\bibinfo  {journal}
  {Royal Society Open Science}\ }\textbf {\bibinfo {volume} {5}},\ \bibinfo
  {pages} {171628} (\bibinfo {year} {2018})}\BibitemShut {NoStop}%
\bibitem [{\citenamefont {Omelyan}\ \emph {et~al.}(2002)\citenamefont
  {Omelyan}, \citenamefont {Mryglod},\ and\ \citenamefont
  {Folk}}]{Omelyan2002OptimizedSystems}%
  \BibitemOpen
  \bibfield  {author} {\bibinfo {author} {\bibfnamefont {I.~P.}\ \bibnamefont
  {Omelyan}}, \bibinfo {author} {\bibfnamefont {I.~M.}\ \bibnamefont
  {Mryglod}}, \ and\ \bibinfo {author} {\bibfnamefont {R.}~\bibnamefont
  {Folk}},\ }\href {\doibase 10.1016/S0010-4655(02)00451-4} {\bibfield
  {journal} {\bibinfo  {journal} {Computer Physics Communications}\ }\textbf
  {\bibinfo {volume} {146}},\ \bibinfo {pages} {188} (\bibinfo {year}
  {2002})}\BibitemShut {NoStop}%
\bibitem [{\citenamefont {Landau}\ \emph {et~al.}(1986)\citenamefont {Landau},
  \citenamefont {Lifshitz}, \citenamefont {Kosevich},\ and\ \citenamefont
  {Pitaevskii}}]{Landau1986TheoryElasticity}%
  \BibitemOpen
  \bibfield  {author} {\bibinfo {author} {\bibfnamefont {L.~D.}\ \bibnamefont
  {Landau}}, \bibinfo {author} {\bibfnamefont {E.~M.}\ \bibnamefont
  {Lifshitz}}, \bibinfo {author} {\bibfnamefont {A.~M.}\ \bibnamefont
  {Kosevich}}, \ and\ \bibinfo {author} {\bibfnamefont {L.~P.}\ \bibnamefont
  {Pitaevskii}},\ }\href@noop {} {\emph {\bibinfo {title} {{Theory of
  elasticity}}}}\ (\bibinfo  {publisher} {Butterworth-Heinemann},\ \bibinfo
  {year} {1986})\ p.\ \bibinfo {pages} {187}\BibitemShut {NoStop}%
\bibitem [{\citenamefont {Audoly}\ \emph {et~al.}(2013)\citenamefont {Audoly},
  \citenamefont {Clauvelin}, \citenamefont {Brun}, \citenamefont {Bergou},
  \citenamefont {Grinspun},\ and\ \citenamefont
  {Wardetzky}}]{Audoly2013AThreads}%
  \BibitemOpen
  \bibfield  {author} {\bibinfo {author} {\bibfnamefont {B.}~\bibnamefont
  {Audoly}}, \bibinfo {author} {\bibfnamefont {N.}~\bibnamefont {Clauvelin}},
  \bibinfo {author} {\bibfnamefont {P.~T.}\ \bibnamefont {Brun}}, \bibinfo
  {author} {\bibfnamefont {M.}~\bibnamefont {Bergou}}, \bibinfo {author}
  {\bibfnamefont {E.}~\bibnamefont {Grinspun}}, \ and\ \bibinfo {author}
  {\bibfnamefont {M.}~\bibnamefont {Wardetzky}},\ }\href {\doibase
  10.1016/j.jcp.2013.06.034} {\bibfield  {journal} {\bibinfo  {journal}
  {Journal of Computational Physics}\ }\textbf {\bibinfo {volume} {253}},\
  \bibinfo {pages} {18} (\bibinfo {year} {2013})}\BibitemShut {NoStop}%
\bibitem [{\citenamefont {Schramm}\ \emph
  {et~al.}(2017{\natexlab{b}})\citenamefont {Schramm}, \citenamefont {Hocky},
  \citenamefont {Voth}, \citenamefont {Blanchoin}, \citenamefont {Martiel},\
  and\ \citenamefont {De~La~Cruz}}]{Schramm2017ActinDissociationb}%
  \BibitemOpen
  \bibfield  {author} {\bibinfo {author} {\bibfnamefont {A.~C.}\ \bibnamefont
  {Schramm}}, \bibinfo {author} {\bibfnamefont {G.~M.}\ \bibnamefont {Hocky}},
  \bibinfo {author} {\bibfnamefont {G.~A.}\ \bibnamefont {Voth}}, \bibinfo
  {author} {\bibfnamefont {L.}~\bibnamefont {Blanchoin}}, \bibinfo {author}
  {\bibfnamefont {J.~L.}\ \bibnamefont {Martiel}}, \ and\ \bibinfo {author}
  {\bibfnamefont {E.~M.}\ \bibnamefont {De~La~Cruz}},\ }\href {\doibase
  10.1016/j.bpj.2017.05.016} {\bibfield  {journal} {\bibinfo  {journal}
  {Biophysical Journal}\ }\textbf {\bibinfo {volume} {112}},\ \bibinfo {pages}
  {2624} (\bibinfo {year} {2017}{\natexlab{b}})}\BibitemShut {NoStop}%
\bibitem [{\citenamefont {Schramm}\ \emph
  {et~al.}(2019{\natexlab{b}})\citenamefont {Schramm}, \citenamefont {Hocky},
  \citenamefont {Voth}, \citenamefont {Martiel},\ and\ \citenamefont
  {De~La~Cruz}}]{Schramm2019PlasticFilamentsb}%
  \BibitemOpen
  \bibfield  {author} {\bibinfo {author} {\bibfnamefont {A.~C.}\ \bibnamefont
  {Schramm}}, \bibinfo {author} {\bibfnamefont {G.~M.}\ \bibnamefont {Hocky}},
  \bibinfo {author} {\bibfnamefont {G.~A.}\ \bibnamefont {Voth}}, \bibinfo
  {author} {\bibfnamefont {J.~L.}\ \bibnamefont {Martiel}}, \ and\ \bibinfo
  {author} {\bibfnamefont {E.~M.}\ \bibnamefont {De~La~Cruz}},\ }\href
  {\doibase 10.1016/j.bpj.2019.06.018} {\bibfield  {journal} {\bibinfo
  {journal} {Biophysical Journal}\ }\textbf {\bibinfo {volume} {117}},\
  \bibinfo {pages} {453} (\bibinfo {year} {2019}{\natexlab{b}})}\BibitemShut
  {NoStop}%
\bibitem [{\citenamefont {Gere}\ and\ \citenamefont
  {Timoshenko}(1997)}]{gere1997mechanics}%
  \BibitemOpen
  \bibfield  {author} {\bibinfo {author} {\bibfnamefont {J.}~\bibnamefont
  {Gere}}\ and\ \bibinfo {author} {\bibfnamefont {S.}~\bibnamefont
  {Timoshenko}},\ }\href {https://books.google.co.uk/books?id=BKJRAAAAMAAJ}
  {\emph {\bibinfo {title} {Mechanics of Materials}}}\ (\bibinfo  {publisher}
  {PWS Publishing Company},\ \bibinfo {year} {1997})\BibitemShut {NoStop}%
\bibitem [{\citenamefont {Leijnse}\ \emph {et~al.}(2015)\citenamefont
  {Leijnse}, \citenamefont {Oddershede},\ and\ \citenamefont
  {Bendix}}]{Leijnse2015HelicalTraction}%
  \BibitemOpen
  \bibfield  {author} {\bibinfo {author} {\bibfnamefont {N.}~\bibnamefont
  {Leijnse}}, \bibinfo {author} {\bibfnamefont {L.~B.}\ \bibnamefont
  {Oddershede}}, \ and\ \bibinfo {author} {\bibfnamefont {P.~M.}\ \bibnamefont
  {Bendix}},\ }\href {\doibase 10.1073/PNAS.1411761112/-/DCSUPPLEMENTAL}
  {\bibfield  {journal} {\bibinfo  {journal} {Proceedings of the National
  Academy of Sciences of the United States of America}\ }\textbf {\bibinfo
  {volume} {112}},\ \bibinfo {pages} {136} (\bibinfo {year}
  {2015})}\BibitemShut {NoStop}%
\bibitem [{\citenamefont {Michaels}\ \emph {et~al.}(2020)\citenamefont
  {Michaels}, \citenamefont {Memet},\ and\ \citenamefont
  {Mahadevan}}]{Michaels2020MechanicalMorphology}%
  \BibitemOpen
  \bibfield  {author} {\bibinfo {author} {\bibfnamefont {T.~C.}\ \bibnamefont
  {Michaels}}, \bibinfo {author} {\bibfnamefont {E.}~\bibnamefont {Memet}}, \
  and\ \bibinfo {author} {\bibfnamefont {L.}~\bibnamefont {Mahadevan}},\ }\href
  {\doibase 10.1039/D0SM01145B} {\bibfield  {journal} {\bibinfo  {journal}
  {Soft Matter}\ }\textbf {\bibinfo {volume} {16}},\ \bibinfo {pages} {9306}
  (\bibinfo {year} {2020})}\BibitemShut {NoStop}%
\bibitem [{\citenamefont {Bergou}\ \emph {et~al.}(2008)\citenamefont {Bergou},
  \citenamefont {Wardetzky}, \citenamefont {Robinson}, \citenamefont {Audoly},\
  and\ \citenamefont {Grinspun}}]{Bergou2008DiscreteRods}%
  \BibitemOpen
  \bibfield  {author} {\bibinfo {author} {\bibfnamefont {M.}~\bibnamefont
  {Bergou}}, \bibinfo {author} {\bibfnamefont {M.}~\bibnamefont {Wardetzky}},
  \bibinfo {author} {\bibfnamefont {S.}~\bibnamefont {Robinson}}, \bibinfo
  {author} {\bibfnamefont {B.}~\bibnamefont {Audoly}}, \ and\ \bibinfo {author}
  {\bibfnamefont {E.}~\bibnamefont {Grinspun}},\ }\href {\doibase
  10.1145/1360612.1360662} {\bibfield  {journal} {\bibinfo  {journal} {ACM
  Transactions on Graphics}\ }\textbf {\bibinfo {volume} {27}} (\bibinfo {year}
  {2008}),\ 10.1145/1360612.1360662}\BibitemShut {NoStop}%
\bibitem [{\citenamefont {Bergou}\ \emph {et~al.}(2010)\citenamefont {Bergou},
  \citenamefont {Audoly}, \citenamefont {Vouga}, \citenamefont {Wardetzky},\
  and\ \citenamefont {Grinspun}}]{Bergou2010DiscreteThreads}%
  \BibitemOpen
  \bibfield  {author} {\bibinfo {author} {\bibfnamefont {M.}~\bibnamefont
  {Bergou}}, \bibinfo {author} {\bibfnamefont {B.}~\bibnamefont {Audoly}},
  \bibinfo {author} {\bibfnamefont {E.}~\bibnamefont {Vouga}}, \bibinfo
  {author} {\bibfnamefont {M.}~\bibnamefont {Wardetzky}}, \ and\ \bibinfo
  {author} {\bibfnamefont {E.}~\bibnamefont {Grinspun}},\ }in\ \href {\doibase
  10.1145/1778765.1778853} {\emph {\bibinfo {booktitle} {ACM SIGGRAPH 2010
  Papers, SIGGRAPH 2010}}},\ Vol.~\bibinfo {volume} {29}\ (\bibinfo {year}
  {2010})\BibitemShut {NoStop}%
\bibitem [{\citenamefont {Yamaoka}\ and\ \citenamefont
  {Adachi}(2010)}]{Yamaoka2010CouplingAxis}%
  \BibitemOpen
  \bibfield  {author} {\bibinfo {author} {\bibfnamefont {H.}~\bibnamefont
  {Yamaoka}}\ and\ \bibinfo {author} {\bibfnamefont {T.}~\bibnamefont
  {Adachi}},\ }\href {\doibase 10.1016/j.ijmecsci.2009.09.038} {\bibfield
  {journal} {\bibinfo  {journal} {International Journal of Mechanical
  Sciences}\ } (\bibinfo {year} {2010}),\
  10.1016/j.ijmecsci.2009.09.038}\BibitemShut {NoStop}%
\bibitem [{\citenamefont {Maxian}\ \emph {et~al.}(2022)\citenamefont {Maxian},
  \citenamefont {Sprinkle}, \citenamefont {Peskin},\ and\ \citenamefont
  {Donev}}]{maxian2022hydrodynamics}%
  \BibitemOpen
  \bibfield  {author} {\bibinfo {author} {\bibfnamefont {O.}~\bibnamefont
  {Maxian}}, \bibinfo {author} {\bibfnamefont {B.}~\bibnamefont {Sprinkle}},
  \bibinfo {author} {\bibfnamefont {C.~S.}\ \bibnamefont {Peskin}}, \ and\
  \bibinfo {author} {\bibfnamefont {A.}~\bibnamefont {Donev}},\ }\href@noop {}
  {\bibfield  {journal} {\bibinfo  {journal} {arXiv preprint arXiv:2201.04187}\
  } (\bibinfo {year} {2022})}\BibitemShut {NoStop}%
\bibitem [{\citenamefont {Nedelec}\ and\ \citenamefont
  {Foethke}(2007)}]{Nedelec2007CollectiveFibers}%
  \BibitemOpen
  \bibfield  {author} {\bibinfo {author} {\bibfnamefont {F.}~\bibnamefont
  {Nedelec}}\ and\ \bibinfo {author} {\bibfnamefont {D.}~\bibnamefont
  {Foethke}},\ }\href {\doibase 10.1088/1367-2630/9/11/427} {\bibfield
  {journal} {\bibinfo  {journal} {New Journal of Physics}\ }\textbf {\bibinfo
  {volume} {9}},\ \bibinfo {pages} {427} (\bibinfo {year} {2007})}\BibitemShut
  {NoStop}%
\bibitem [{\citenamefont {Freedman}\ \emph {et~al.}(2017)\citenamefont
  {Freedman}, \citenamefont {Banerjee}, \citenamefont {Hocky},\ and\
  \citenamefont {Dinner}}]{freedman2017versatile}%
  \BibitemOpen
  \bibfield  {author} {\bibinfo {author} {\bibfnamefont {S.~L.}\ \bibnamefont
  {Freedman}}, \bibinfo {author} {\bibfnamefont {S.}~\bibnamefont {Banerjee}},
  \bibinfo {author} {\bibfnamefont {G.~M.}\ \bibnamefont {Hocky}}, \ and\
  \bibinfo {author} {\bibfnamefont {A.~R.}\ \bibnamefont {Dinner}},\
  }\href@noop {} {\bibfield  {journal} {\bibinfo  {journal} {Biophysical
  journal}\ }\textbf {\bibinfo {volume} {113}},\ \bibinfo {pages} {448}
  (\bibinfo {year} {2017})}\BibitemShut {NoStop}%
\bibitem [{\citenamefont {Galkin}\ \emph {et~al.}(2015)\citenamefont {Galkin},
  \citenamefont {Orlova}, \citenamefont {Vos}, \citenamefont {Schr{\"{o}}der},\
  and\ \citenamefont {Egelman}}]{Galkin2015Near-AtomicF-Actin}%
  \BibitemOpen
  \bibfield  {author} {\bibinfo {author} {\bibfnamefont {V.~E.}\ \bibnamefont
  {Galkin}}, \bibinfo {author} {\bibfnamefont {A.}~\bibnamefont {Orlova}},
  \bibinfo {author} {\bibfnamefont {M.~R.}\ \bibnamefont {Vos}}, \bibinfo
  {author} {\bibfnamefont {G.~F.}\ \bibnamefont {Schr{\"{o}}der}}, \ and\
  \bibinfo {author} {\bibfnamefont {E.~H.}\ \bibnamefont {Egelman}},\ }\href
  {\doibase 10.1016/j.str.2014.11.006} {\bibfield  {journal} {\bibinfo
  {journal} {Structure}\ }\textbf {\bibinfo {volume} {23}},\ \bibinfo {pages}
  {173} (\bibinfo {year} {2015})}\BibitemShut {NoStop}%
\bibitem [{\citenamefont {Fan}\ \emph {et~al.}(2013)\citenamefont {Fan},
  \citenamefont {Saunders}, \citenamefont {Haddadian}, \citenamefont {Freed},
  \citenamefont {De}, \citenamefont {Cruz},\ and\ \citenamefont
  {Voth}}]{Fan2013MolecularMechanics}%
  \BibitemOpen
  \bibfield  {author} {\bibinfo {author} {\bibfnamefont {J.}~\bibnamefont
  {Fan}}, \bibinfo {author} {\bibfnamefont {M.~G.}\ \bibnamefont {Saunders}},
  \bibinfo {author} {\bibfnamefont {E.~J.}\ \bibnamefont {Haddadian}}, \bibinfo
  {author} {\bibfnamefont {K.~F.}\ \bibnamefont {Freed}}, \bibinfo {author}
  {\bibfnamefont {E.~M.}\ \bibnamefont {De}}, \bibinfo {author} {\bibfnamefont
  {L.}~\bibnamefont {Cruz}}, \ and\ \bibinfo {author} {\bibfnamefont {G.~A.}\
  \bibnamefont {Voth}},\ }\href {\doibase 10.1016/j.jmb.2013.01.020} {\bibfield
   {journal} {\bibinfo  {journal} {Journal of Molecular Biology}\ } (\bibinfo
  {year} {2013}),\ 10.1016/j.jmb.2013.01.020}\BibitemShut {NoStop}%
\bibitem [{\citenamefont {Grazi}()}]{Grazi1997WhatFilament}%
  \BibitemOpen
  \bibfield  {author} {\bibinfo {author} {\bibfnamefont {E.}~\bibnamefont
  {Grazi}},\ }\href {\doibase 10.1016/S0014-5793(97)00214-7} {\bibfield
  {journal} {\bibinfo  {journal} {FEBS Letters}\ }\textbf {\bibinfo {volume}
  {405}},\ \bibinfo {pages} {249}}\BibitemShut {NoStop}%
\bibitem [{\citenamefont {Tseng}\ \emph {et~al.}(2002)\citenamefont {Tseng},
  \citenamefont {Schafer}, \citenamefont {Almo},\ and\ \citenamefont
  {Wirtz}}]{Tseng2002FunctionalProteins}%
  \BibitemOpen
  \bibfield  {author} {\bibinfo {author} {\bibfnamefont {Y.}~\bibnamefont
  {Tseng}}, \bibinfo {author} {\bibfnamefont {B.~W.}\ \bibnamefont {Schafer}},
  \bibinfo {author} {\bibfnamefont {S.~C.}\ \bibnamefont {Almo}}, \ and\
  \bibinfo {author} {\bibfnamefont {D.}~\bibnamefont {Wirtz}},\ }\href
  {\doibase 10.1074/jbc.M202609200} {\bibfield  {journal} {\bibinfo  {journal}
  {Journal of Biological Chemistry}\ }\textbf {\bibinfo {volume} {277}},\
  \bibinfo {pages} {25609} (\bibinfo {year} {2002})}\BibitemShut {NoStop}%
\bibitem [{\citenamefont {Kojima}\ \emph {et~al.}(1994)\citenamefont {Kojima},
  \citenamefont {Ishijima},\ and\ \citenamefont
  {Yanagida}}]{Kojima1994DirectNanomanipulation}%
  \BibitemOpen
  \bibfield  {author} {\bibinfo {author} {\bibfnamefont {H.}~\bibnamefont
  {Kojima}}, \bibinfo {author} {\bibfnamefont {A.}~\bibnamefont {Ishijima}}, \
  and\ \bibinfo {author} {\bibfnamefont {T.}~\bibnamefont {Yanagida}},\ }\href
  {\doibase 10.1073/pnas.91.26.12962} {\bibfield  {journal} {\bibinfo
  {journal} {Proceedings of the National Academy of Sciences of the United
  States of America}\ }\textbf {\bibinfo {volume} {91}},\ \bibinfo {pages}
  {12962} (\bibinfo {year} {1994})}\BibitemShut {NoStop}%
\bibitem [{\citenamefont {McCullough}\ \emph {et~al.}(2008)\citenamefont
  {McCullough}, \citenamefont {Blanchoin}, \citenamefont {Martiel},\ and\
  \citenamefont {De~La~Cruz}}]{McCullough2008CofilinMechanicsb}%
  \BibitemOpen
  \bibfield  {author} {\bibinfo {author} {\bibfnamefont {B.~R.}\ \bibnamefont
  {McCullough}}, \bibinfo {author} {\bibfnamefont {L.}~\bibnamefont
  {Blanchoin}}, \bibinfo {author} {\bibfnamefont {J.~L.}\ \bibnamefont
  {Martiel}}, \ and\ \bibinfo {author} {\bibfnamefont {E.~M.}\ \bibnamefont
  {De~La~Cruz}},\ }\href {\doibase 10.1016/J.JMB.2008.05.055} {\bibfield
  {journal} {\bibinfo  {journal} {Journal of Molecular Biology}\ }\textbf
  {\bibinfo {volume} {381}},\ \bibinfo {pages} {550} (\bibinfo {year}
  {2008})}\BibitemShut {NoStop}%
\bibitem [{\citenamefont {Isambert}\ \emph {et~al.}(1995)\citenamefont
  {Isambert}, \citenamefont {Venier}, \citenamefont {Maggs}, \citenamefont
  {Fattoum}, \citenamefont {Kassab}, \citenamefont {Pantaloni},\ and\
  \citenamefont {Carlier}}]{Isambert1995Flexibilityproteins}%
  \BibitemOpen
  \bibfield  {author} {\bibinfo {author} {\bibfnamefont {H.}~\bibnamefont
  {Isambert}}, \bibinfo {author} {\bibfnamefont {P.}~\bibnamefont {Venier}},
  \bibinfo {author} {\bibfnamefont {A.~C.}\ \bibnamefont {Maggs}}, \bibinfo
  {author} {\bibfnamefont {A.}~\bibnamefont {Fattoum}}, \bibinfo {author}
  {\bibfnamefont {R.}~\bibnamefont {Kassab}}, \bibinfo {author} {\bibfnamefont
  {D.}~\bibnamefont {Pantaloni}}, \ and\ \bibinfo {author} {\bibfnamefont
  {M.~F.}\ \bibnamefont {Carlier}},\ }\href {\doibase 10.1074/jbc.270.19.11437}
  {\bibfield  {journal} {\bibinfo  {journal} {Journal of Biological Chemistry}\
  }\textbf {\bibinfo {volume} {270}},\ \bibinfo {pages} {11437} (\bibinfo
  {year} {1995})}\BibitemShut {NoStop}%
\bibitem [{\citenamefont {Ott}\ \emph {et~al.}(1993)\citenamefont {Ott},
  \citenamefont {Magnasco}, \citenamefont {Simon},\ and\ \citenamefont
  {Libchaber}}]{Ott1993MeasurementMicroscopy}%
  \BibitemOpen
  \bibfield  {author} {\bibinfo {author} {\bibfnamefont {A.}~\bibnamefont
  {Ott}}, \bibinfo {author} {\bibfnamefont {M.}~\bibnamefont {Magnasco}},
  \bibinfo {author} {\bibfnamefont {A.}~\bibnamefont {Simon}}, \ and\ \bibinfo
  {author} {\bibfnamefont {A.}~\bibnamefont {Libchaber}},\ }\href {\doibase
  10.1103/PhysRevE.48.R1642} {\bibfield  {journal} {\bibinfo  {journal}
  {Physical Review E}\ }\textbf {\bibinfo {volume} {48}},\ \bibinfo {pages}
  {R1642} (\bibinfo {year} {1993})}\BibitemShut {NoStop}%
\bibitem [{\citenamefont {Ling}\ \emph {et~al.}(2020)\citenamefont {Ling},
  \citenamefont {Wei}, \citenamefont {Wang}, \citenamefont {Yang},
  \citenamefont {Qu},\ and\ \citenamefont
  {Fang}}]{Ling2020ExperimentallyMetamaterials}%
  \BibitemOpen
  \bibfield  {author} {\bibinfo {author} {\bibfnamefont {B.}~\bibnamefont
  {Ling}}, \bibinfo {author} {\bibfnamefont {K.}~\bibnamefont {Wei}}, \bibinfo
  {author} {\bibfnamefont {Z.}~\bibnamefont {Wang}}, \bibinfo {author}
  {\bibfnamefont {X.}~\bibnamefont {Yang}}, \bibinfo {author} {\bibfnamefont
  {Z.}~\bibnamefont {Qu}}, \ and\ \bibinfo {author} {\bibfnamefont
  {D.}~\bibnamefont {Fang}},\ }\href {\doibase 10.1016/J.IJMECSCI.2020.105466}
  {\bibfield  {journal} {\bibinfo  {journal} {International Journal of
  Mechanical Sciences}\ }\textbf {\bibinfo {volume} {173}},\ \bibinfo {pages}
  {105466} (\bibinfo {year} {2020})}\BibitemShut {NoStop}%
\end{thebibliography}
%
\end{document}